# Ultrafast photochemistry and electron-diffraction spectra in n→(3s) Rydberg excited cyclobutanone resolved at the multireference perturbative level.

V. K. Jaiswal,[†,1] F. Montorsi,[†,1] F. Aleotti,[†,1] F. Segatta,[1] Daniel Keefer,[2] Shaul Mukamel,[3] A. Nenov*,[1] I. Conti*,[1] M. Garavelli*,[1]

[1]*Dipartimento di Chimica industriale "Toso Montanari", Università di Bologna, Viale del Risorgimento 4, 40136 Bologna, Italy*

[2]*Molecular Spectroscopy Department, Max Planck Institute for Polymer Research, Mainz 55128, Germany*

[3]*Department of Chemistry and Department of Physics & Astronomy, University of California Irvine, Irvine, California 92697, United States*

[†] These authors have contributed equally*(Irene Conti):  irene.conti@unibo.it, (Artur Nenov): artur.nenov@unibo.it,

(Marco Garavelli): marco.garavelli@unibo.it

We study the ultrafast time evolution of cyclobutanone excited to singlet n→Rydberg state through XMS-CASPT2 non-adiabatic surface-hopping simulations. These dynamics predict relaxation to ground-state with a timescale of 822 ± 45 fs with minimal involvement of triplets. The major relaxation path to the ground-state involves a three-state degeneracy region and leads to variety of fragmented photoproducts. We simulate the resulting time-resolved electron-diffraction spectra which track the relaxation of the excited state and the formation of various photoproducts in the ground-state.

## I. INTRODUCTION

The advent of modern ultrafast spectroscopic techniques have enabled the recording of time-resolved photoinduced processes with unprecedented temporal and spatial resolution.[1–4] These spectroscopic techniques have been used to  track complex photoinduced processes involving. the interplay of various structural, electronic, and environmental factors in systems ranging from small isolated molecules in gas-phase to biological systems embedded in physiological environments.[5–8] Thus, computer simulations of light-induced molecular dynamics have become indispensable tools to complement these novel experiments and decode the underlying mechanistic processes hidden in experimental signals. The power of theory has been demonstrated in seminal works where they have helped to unveil the experimentally recorded photoinduced mechanistic pathways with atomistic resolution.[8–12]

At this point one could dare to claim that theoretical modeling has matured enough not only to explain but rather to predict experimental outcomes. In this spirit, a gauntlet has been thrown to the theoretical community to model the ultrafast photochemistry of cyclobutanone and compare against ultrafast electron diffraction experiments to be undertaken at the SLAC Megaelectronvolt facility, thus putting the above statement under scrutiny. The unprecedented spatio-temporal resolution provided by this novel experiment is particularly sensitive to changes in the inter-nuclear distances, and can capture a movie of actual molecule as it undergoes structural deformations along excited state deactivation pathway.[13–15] In case of cyclobutanone the photophysical pathways can lead to various fragmented photoproducts resulting in drastic changes in distribution of inter-nuclear distances which are directly mapped in these experimental electron-diffraction signals. This makes it possible to measure the timescales of dynamics which can be used as a benchmark to assess the quality of dynamical simulations.

The photophysics of cyclobutanone has been a subject of active experimental and theoretical investigations.[16–31] The experimental absorption spectra characterize the two lowest singlet states as a $n\pi^*$ and n→(3s)Rydberg excitation.[22,25,26] The photoinduced processes following excitation to the $n\pi^*$ state have been comprehensively explored. The photolysis of cyclobutanone excited to this lowest excited state yields C3 (cyclopropane) and C2 (ethene + ketene) fragmented species as major photoproducts.[16,17] Recent ultrafast experimental works have focused on resolving the timescale of these photochemical processes. Kao[21] et al. reported. a sub-picosecond timescale for the cleavage of C-C bonds and formation of ketene containing photoproducts and interpreted the fast timescales due to a high ring-strain and a low transition state of 29 kJ/mol (0.3 eV) needed to overcome for this process. Xia et al.[23] made an extensive analysis of the reactive potential energy surface of the $n\pi^*$ state relevant for ring-opening reactions at MS-CASPT2 level and resolve a barrier of 6.7 kcal/mol (0.29 eV) for this reaction in agreement with previous experimental works.[19,21,24] Liu et al.[27] augmented the theoretical picture with dynamical simulations on the $n\pi^*$ state using Ab-initio Multiple Spawning and resolved a timescale of 176 fs for cleavage of α C-C bonds and 484 fs for the internal conversion to the ground state.

In comparison, the photoinduced evolution post excitation to the 3s Rydberg state is still lacking. Kuhlman et al. studied the dynamics of excitation to the 3s Rydberg state at 200nm using ultrafast experimental techniques and theoretical vibronic coupling models.[18,20,29] They observed a biexponential timescale of 80fs and 740 fs in time-resolved mass-spectrometry and 950fs in time-resolved photoelectron spectroscopy.[20,29] Using a vibronic coupling model they assigned these dynamics to internal conversion from the Rydberg state to the $n\pi^*$ state.

This study strives to resolve the ultrafast dynamics of cyclobutanone excited to the 3s Rydberg state at 200 nm. We employ multiconfigurational wavefunctions including dynamical correlation as implemented in the XMS-CASPT2 protocol.[32–34] This approach allows for quantitively accurate energy surfaces and has the flexibility required to describe bond-breaking events. The dynamics are executed within a mixed quantum-classical formalism employing surface-hopping, a method extensively utilized to interpret ultrafast experiments with notable success.[35–37]

## II. COMPUTATIONAL DETAILS

A multiconfigurational wavefunction is paramount for the correct description of the photoinduced relaxation and formation of various photoproducts which might involve bond-breaking. CASPT2 based methods were therefore employed for optimizations of critical structures, reaction paths and non-adiabatic dynamics.

### A. ELECTRONIC STRUCTURE BENCHMARKING

The ground-state was optimized at Möller-Plesset second order perturbation theory (MP2) level employing the 6-31++G* basis set, which includes diffuse functions necessary to describe excitations into Rydberg orbitals. Vertical energies were computed using the XMS variant of CASPT2 method (XMS-CASPT2)[32–34], using a state-averaged CASSCF wavefunction with a |12,12| active-space employed in previous studies[27] as a benchmark to assess the suitability of a reduced |8,8| active-space employed in this work. The orbitals included in these active-spaces are shown in Figure S1. The lowest three singlet/triplet states were obtained in two separate state-averaged CASSCF calculations. The spin-orbit coupling (SOC) between the XMS-CASPT2 states were computed *a posteriori* through a spin-orbit Hamiltonian approximated by a one-electron effective Hamiltonian using Atomic Mean Field Approximation.[38] An imaginary shift[39] of 0.2 was employed to avoid intruder states in CASPT2 computations. All XMS-CASPT2 computations were done using zero IPEA shift.

### B. NON-ADIABATIC DYNAMICS

The nonadiabatic mixed quantum-classical dynamics at XMS-CASPT2 level with an |8,8| active-space and CASSCF level with |12,12| active-space were performed with a fewest switches Tully surface-hopping algorithm[40] with Tully-Hammes-Schiffer modification[41] (THS) including decoherence corrections.[42] The THS scheme is realized by computing directly the time-derivative couplings by means of wave function overlaps at consecutive geometries along a trajectory, thus omitting the need to compute spatial derivative-couplings. These dynamics were done on an ensemble of 100 structures generated through Wigner sampling using harmonic frequencies on top of the optimized ground-state geometry at MP2 level. Seven normal modes were frozen in this sampling including all the CH modes with frequencies larger than 3000 cm$^{-1}$ and the normal mode with the smallest frequency lower than 100 cm$^{-1}$. A time-step of 1 fs was utilized in surface-hopping dynamics. This comparatively large timestep is enabled by our choice of hopping scheme, namely THS, as wave function overlaps allow to trace non-adiabatic behavior over a

broader spatial region compared to spatial derivative couplings whose spiked nature encompasses highly local information about the nonadiabaticity. The nuclear and electronic dynamics were propagated using analytical gradients at XMS-CASPT2 level[43] and derivative couplings computed through wavefunction overlaps[44] of the electronic states between successive timesteps. After the hop the velocities were uniformly rescaled along all coordinates. Intersystem crossing was modelled with the spin-diabatic approach described by Cui et al.[45] and approximating the SX-TY spin-orbit coupling as the sum of the moduli of the couplings associated to all spin sublevels (mS=-1,0,1) of the given triplet state Y. Non-adiabatic surface-hopping molecular dynamics including ISC was performed with a developer version of the COBRAMM[46,47] software interfaced with OpenMolcas.[48]

To estimate the errors due to finite size of the ensemble (100 trajectories) bootstrapping methodology was employed.[49] In every bootstrap cycle, a new bootstrap ensemble of 100 trajectories were created by sampling with replacement from the original ensemble. The quantity of interest (e.g., lifetime of a monoexponential fit over average population) was computed for every new bootstrap ensemble. The reported standard deviation refers to 10,000 bootstrapping cycles.

## C. MODELING OF TIME-RESOLVED UED SPECTRA

Ultrafast electron diffraction (UED) signals were simulated within the Independent Atom Model (IAM) utilizing the nuclear configurations extracted at every timestep from the surface-hopping dynamics, as described in reference[15]. The strength of the individual atom approximation in this system has been verified, on some reference molecular structure, by comparing UED signals computed from *ab-initio* electron densities with the IAM predictions. The result of this comparison is reported in the supporting information.

The raw diffraction pattern was then post-processed to extract time dependent modified scattering intensity $\Delta sM(t,q)$ (reported in Figure S7 of the supporting information). Utilizing a sine-transform of the modified diffraction pattern the real space difference pair distribution function (ΔPDF) was obtained. This signal is proportional to the probability of finding an atom pair at a given distance $r$, allowing to extract direct information about structural modifications along the passage of photochemical pathways following molecular photoexcitation[13,14]. To compare the predicted diffraction pattern with its experimental counterpart, the finite time resolution of the experiment was also considered by convoluting the signal with a Gaussian envelope function characterized by a time resolution of 150 fs (FWHM). More details about the simulation protocol are reported in the supporting information.

## III. RESULTS and DISCUSSION

## A. VERTICAL EXCITATION ENERGIES

In Table I we report the CASSCF and XMS-CASPT2 vertical excitation energies of the lowest singlets and triplets at Franck-Condon geometry employing the |12,12| and |8,8| active-spaces. The lowest two singlet excited states are respectively, a dark n→π* state ($S_1$) at ~ 4.22 eV and a n→Rydberg state ($S_2$) at ~6.30 eV at XMS-CASPT2 level. There are three triplet states energetically below the singlet $S_2$ excitation at Franck-Condon geometry. The lowest triplet state T1 of n→π* character is ~0.25 eV below the $S_1$ state. The other two triplets - of π→π* ($T_2$) and n→Rydberg ($T_3$) character, are energetically near-degenerate with the $S_2$. The importance of including dynamical correlation on top of CASSCF wavefunction is underlined while considering the energetics of the π→π* triplet, whose energy is significantly underestimated at CASSCF level. The comparison of the two active spaces energetics at the FC and along the main reaction coordinate (Figure S11) supports the choice of the smaller one for the dynamics.

## B. NON-ADIABATIC DYNAMICS

In Figure 1(a) we show the time-resolved population of the singlet and triplets along 2ps of non-adiabatic dynamics initialized on the bright $S_2$ state at Franck-Condon region. 100 surface-hopping trajectories were run at XMS-CASPT2 level including three lowest singlet and triplet states, of which 87 returned to ground-state within 2ps, with the rest remained trapped in $S_2$ state. A monoexponential model of the time-resolved electronic population of $S_2$ gives a lifetime of $822 \pm 45$ fs. The $S_1$ state is transiently populated in the dynamics but no significant population is accumulated in this state. After return to ground-state various photoproducts are formed whose distribution is given in Table II. The major photoproducts formed are cyclopropane+CO and ethene+ketene with almost 40% yield each. Minor products with yields of about 5% are propylene as well as reformed cyclobutanone. Our dynamics have been performed by inclusion of three lowest triplets with a spin-diabatic formalism. However, due to ultrafast timescale of ground-state population, the dynamics show negligible involvement of triplet states in photorelaxation.

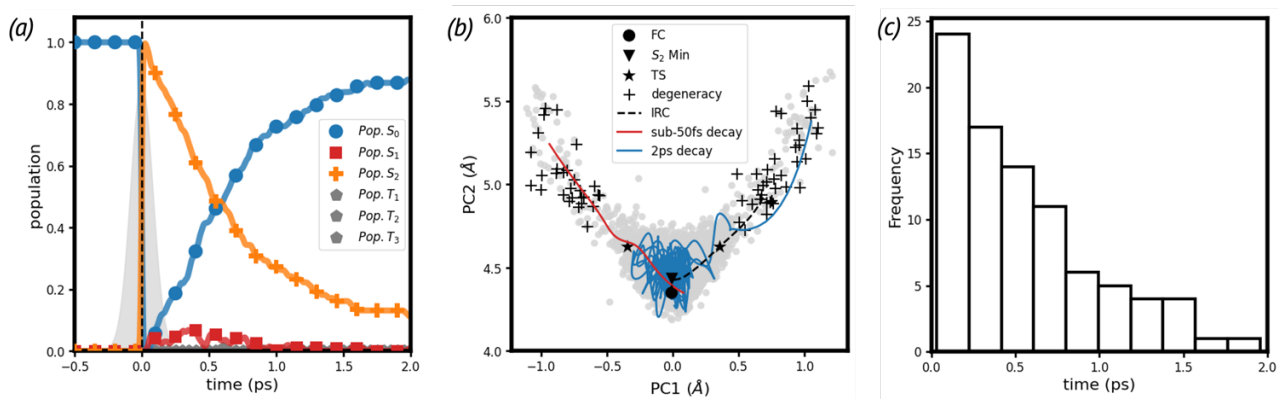

figure 1: (a) Time dependent population of the three lowest singlet and triplet adiabatic states computed through average occupation numbers in the ensemble of trajectories. The occupations obtained from the non-adiabatic dynamics have been further convoluted with a gaussian envelope of 80fs FWHM to mimic the experimental time resolution; (b) Projection of the conformational space spanned by $S_2$ PES from the time of excitation till the arrival to 3-state degeneracy on the two largest principal components. The principal components are obtained by a Singular Value Decomposition of the internal C-C and C-O distances of the molecular structures. Grey dots in background are the structures revealing the span of $S_2$ PES. The critical structures along the reaction pathway are labelled. Two trajectories at the extremes of sub-50fs and ~2ps timescales for $S_2$ decay are shown. (c) Distribution of times needed to arrive at the 3-state degeneracy region.

Almost all trajectories that decay to the ground-state approach a $S_2/S_1/S_0$ 3-state degeneracy region on the $S_2$ PES. In Figure 1(b) we show this first mechanistic step of relaxation on a 2-dimensional projection of the $S_2$ PES. The 2D map is obtained by performing principal component analysis of the trajectories in internal distances between all non-hydrogen atoms from the time of excitation till they first reach this triple degeneracy region and using the two principal coordinates with largest variance capturing most important structural deformations leading to these conformations. The gray dots in background of Figure 1(b) represent all the structures in the dynamical ensemble from initial time till they arrive at the degeneracy region. Two critical regions of the $S_2$-PES along the reaction coordinate to reach the degeneracy region namely the local minima and a transition-state (TS) are also shown. The time to reach this degeneracy region shows a broad distribution from sub-100fs to 2ps as shown in Figure 1(c). Two example trajectories with varying timescales are projected onto the principal coordinates in Figure 1(b). The faster trajectory showing sub-60fs decay to ground-state approaches this CI region in a ballistic way.

Conversely, the other example trajectory decaying at 2ps spends significant amount of time roaming in the vicinity of the n→Rydberg state minimum on the $S_2$ surface.

This local minimum causes a trapping of population, leading to large variance in timescales to reach the $S_2/S_1/S_0$ degeneracy region. As seen in Figure 2 from the XMS-CASPT2 optimized reaction path for this mechanism, a TS which lies 0.19 eV above the local $S_2$-min must be traversed to escape the trap. In this TS (Figure S4) an α C-C bond elongates to 2 Å and the molecule is no more planar with the -CH2- moiety moving out of plane. This results in change of diabatic nature of S2 surface (Figure 2(b-d)) because the bond-elongation and the breaking of planarity radically transform the n-orbital and cause the electron density to be localized in a p-orbital on the -CH2- moiety (Figure 2(d)). The path from this TS to the 3-state degeneracy region is steep and barrierless, with an energy stabilization of ~1.45 eV w.r.t. to n→Rydberg local $S_2$-minima, and involves further changes in diabatic nature of the $S_2$-surface as shown in Figure 2 (d-f). At the midway point from TS to 3-state degeneracy (labelled as point X on the IRC path in Figure 2) further elongation of the α C-C bond changes the nature of the $S_2$ state to a biradical and the $S_0$ state to a zwitterionic closed shell with 2 orbitals localized on -CH2- moiety (Figure 2(e)). From this point X towards the 3-state degeneracy region, the increased mixing of biradical and closed-shell configuration on $S_0/S_2$ leads to a smooth change in diabatic nature of these surfaces. Further deformation along this reaction coordinate can swap the diabatic nature of these surfaces, as evident from the last point on the IRC in Figure 2, where the $S_0$ surface has acquired major biradical character and $S_2$ is majorly a closed-shell zwitterionic structure.

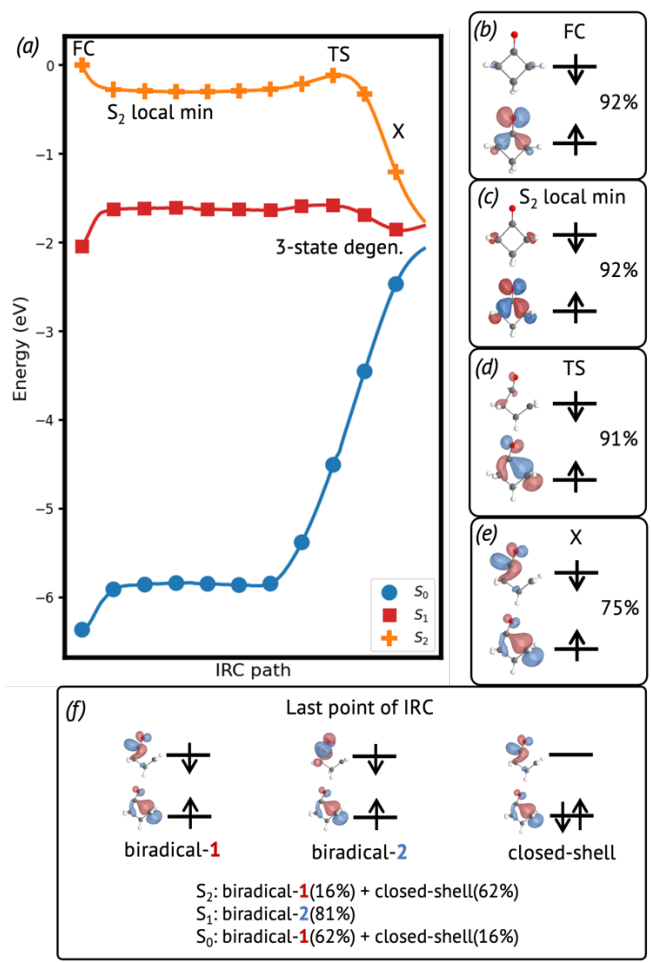

Figure 2: (a) Optimized reaction path from Franck-Condon to the 3-state degeneracy region. (b-e) The major electronic configurations of the $S_2$ state in the XMS-CASPT2 wavefunction in the basis of state-averaged CASSCF orbitals at important critical points of the reaction path along with their respective weights. (f) The three major electronic configurations in the 3 state degeneracy region. The nature of all three adiabatic surfaces $S_0$, $S_1$ and $S_2$ at the last point of IRC is shown in terms of linear combinations of these electronic configurations.

As shown in Figure 2(f), the three major electronic (diabatic) states in this degeneracy region are two biradicals (labelled biradical-**1** and biradical-**2**) involving the localized 2p orbital on -CH2- moiety and two orthogonal π* orbitals on C=O moiety, and a zwitterionic closed shell configuration with both electrons in the p orbital on -CH2-moiety. In this degeneracy region the adiabatic surfaces $S_0/S_1/S_2$ can swap their diabatic nature with ease. The population of the two biradicals leads to the formation of various photoproducts.[50] If the dynamics proceeds on the biradical-**1** (usually the $S_0$ adiabatic state) it leads to formation of both cyclopropane+CO and ethene+ketene. In this case, the formation of photoproducts from the open-ring intermediate happens on the ground $S_0$ adiabatic surface. Instead, if the dynamics proceeds on biradical-**2**, it leads exclusively to formation of ethene + ketene. In this case the formation of photoproducts happens on the $S_1$ surface, leading to excited states localized on two fragments which then return to $S_0$. In Figure 3, we have projected these CI's on their α and β C-C bond-lengths, and color coded them according to the photoproducts being formed. The first kind of CI, where the open-ring intermediate directly hops to $S_0$ from $S_2$, has more pronounced changes in α bond-lengths compared to Franck-Condon. Instead in the other $S_1/S_0$ CI, where the fragmentation into ethene and ketene occurs on $S_1$ surface itself, presents changes in both types of C-C bond-lengths.

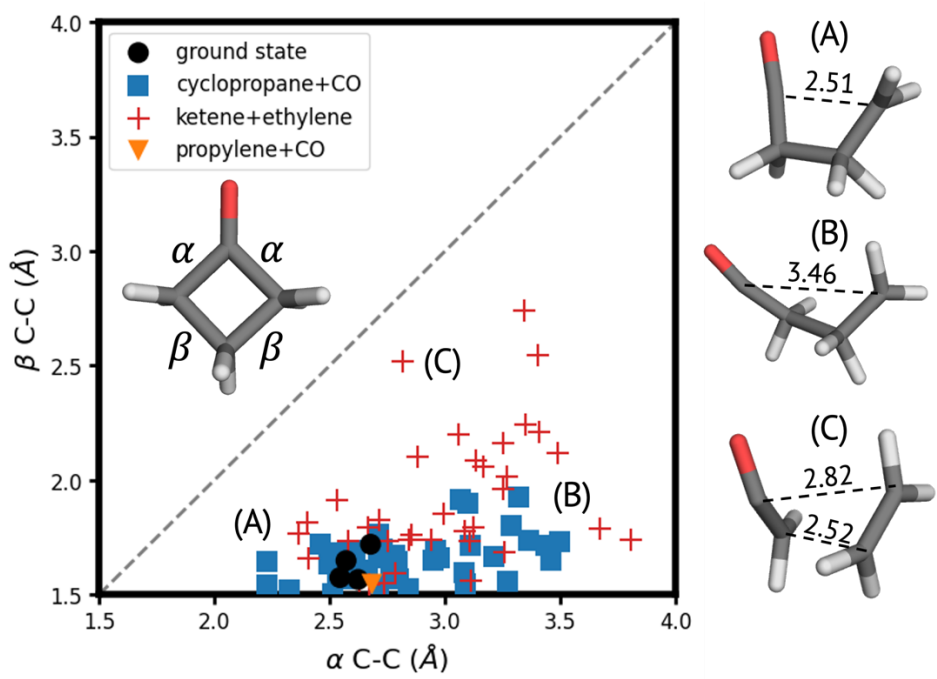

Figure 3: (a) The α and β C-C bond lengths (depicted in the figure inset) of the major conical intersections for hop to $S_0$ observed in the non-adiabatic dynamics.

An analysis of time needed to hop to the $S_0$ surface shows that, once the trajectories escape the trapping region on the $S_2$ PES and cross the TS region, they decay through the CI and photoproduct formation is no more impeded and proceeds ballistically. As shown Figure S12(a) the time taken to hop to the $S_0$ surface is directly correlated with the time taken to reach the degeneracy region. The distribution of time spent in the degeneracy region before hopping to $S_0$ in Figure S12(b) shows that this region is traversed within 300fs with a lifetime of ~40fs given by exponential fitting.

A second kind of mechanism from Franck-Condon is also active where the dynamics approach a $S_2/S_1$ CI instead of 3-state degeneracy region. At this CI, the diabatic character of the $S_2$ and $S_1$ states are the same as Franck-Condon. The minimum energy CI for this mechanism is 0.44 eV above the Franck-Condon, so this mechanism is energetically unfavorable, as evident from only 5/100 trajectories which follow this path. After hopping to $n\pi^*$ $S_1$ state, the dynamics approach a $S_1/S_0$ CI, easily overcoming a barrier of 0.29 eV on the $S_1$ adiabatic surface (Figure S5). Along this path, the diabatic character of $S_1$ changes smoothly from a $n\pi^*$ to a $\sigma\sigma^*$ involving an α C-C bond, promoting cleavage of this particular bond. This open ring intermediate hops to the $S_0$ state and leads to formation of the photoproducts.

We do not observe any significant population of triplet states in our dynamics, save for sporadic hops observed in a couple of trajectories for few femtoseconds, which are just a consequence of random nature of Tully surface hopping algorithm. As shown in Table III, the value of SOC between the singlets and triplets is quite low in this system. The nature of dynamics suggests that the $S_2$ trap region around the local n→Rydberg minima could be a zone of possible ISC, as the $T_2$ state is nearly degenerate with the $S_2$ state here (shown by the $S_2$ and $T_2$ energies documented in Figure S5). While the $S_2$-$T_2$ SOC magnitude is zero at the planar $S_2$ minimum, out of plane vibrations while the system is trapped in this zone could promote ISC through spin-vibronic mechanisms.[51] The reaction path in case of probable $T_2$ population is shown in Figure S6, which predicts a path from $T_2$ to lowest triplet $T_1$ involving a tiny energetic barrier (~ 0.13 eV) and an easy access to the $T_1/S_0$ intersystem crossing with a similar energetic barrier of merely ~0.13 eV on the $T_1$ surface. However, in our simulations the average value of SOC in this trapping region is still quite low (see Table III) and thus we observe no significant ISC.

We also simulated the effect of selective excitation of a narrow energy window on nonadiabatic population dynamics to compare with the experimental setup. The proposed experiment is supposed to excite the experimental vibronic band of the n-> Ryd excitation at 200 nm, which is ~0.2 eV less than the maxima of the experimental absorption band, which we replicate in our setup by selecting geometries with excitation lying within 6 to 6.2 eV. The resulting dynamics are slightly slower compared to excitation of the full Wigner ensemble (Figure S13). A monoexponential fitting of the $S_2$ state gives a slightly larger lifetime of 1089 ± 74 fs. This is expected due to the decreased energy deposited into the system. The quantum yield of the photoproducts (Table S2) too shows similar yields compared dynamics obtained by excitation of the full Wigner ensemble.

## C. TIME-RESOLVED UED SPECTRA

Figure 4(a) displays the real space pair distribution function, $PDF(r)$, of cyclobutanone in its ground state. In this spectrum one recognizes three major features, located at 1.5, 2.3 and 3.2 Å, and respectively associated to 1-2 ($D_1$), 1-3 ($D_2$) and 1-4 ($D_3$) C-C and C-O equilibrium distances.

Since the focus of the UED experiment is on the modification of the electron density (ED) signal upon molecular photoexcitation, the reported signal at a given time delay t (between the pump and the probe pulses) is the difference between the signal of the photoexcited sample and that of the unexcited sample. This leads to a differential time dependent $\Delta PDF(t,r)$ patter which encodes the changes of the molecular structure that follows the photoexcitation. Negative $\Delta PDF(t,r)$ signals are related to the loss of a given equilibrium distance due to bond elongation or breaking[15]. These are typically accompanied by the appearance of corresponding gain (positive) signals, which track the bond distance that is elongating/breaking.

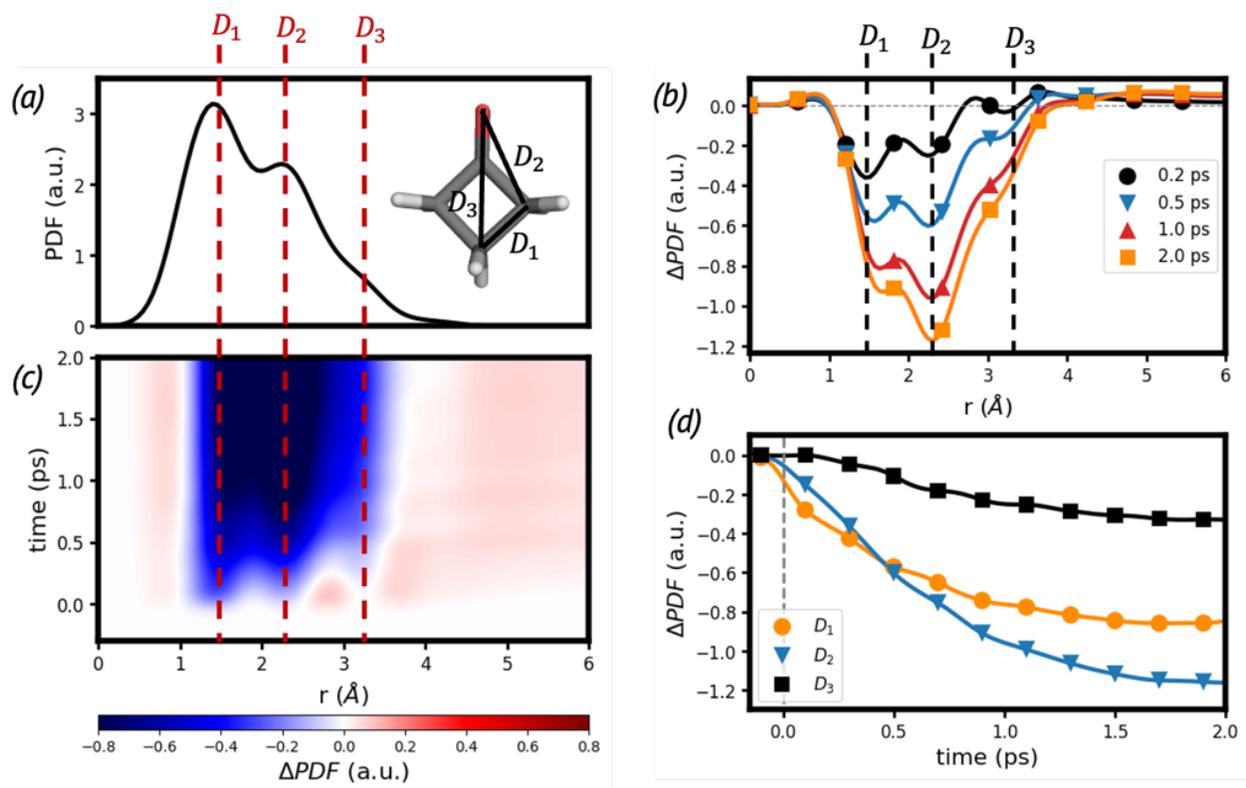

Figure 4: Predicted time dependent $\Delta PDF$ for the cyclobutanone photofragmentation. The static PDF of the unexcited cyclobutanone (a) is here compared with the time dependent $\Delta PDF$ (c). The characteristic

1-2, 1-3 and 1-4 equilibrium distances observed in the ground state cyclobutanone molecule are depicted in the inset of (a) and labeled as $D_1$, $D_2$ and $D_3$. These pair distances are also identified on both (a) and (c) by dashed red lines. Different energy traces of the total $\Delta PDF$ signals are reported in (b) for increasing times from 0.2 to 2.0 ps while the time traces of the characteristic $D_1$, $D_2$ and $D_3$ pair distances are depicted in panel (d).

The predicted $\Delta PDF(t,r)$, reported in Figure 4(c) displays an intense loss of $D_1$, $D_2$ and $D_3$ equilibrium distances. These features are accompanied by both: a weak gain signal around 0.6 Å and a weak and extremely broad gain occupying the whole region of distances greater than $D_3$. The loss of equilibrium distances reflects the breaking of two C-C bonds, and can be used to monitor (in our simulations as well as in the experiment) the occurrence of ring fragmentation. This negative signal builds up in time as displayed by the energy and time traces of Figure 4 (b) and (d) and encodes the timescale of the cyclobutanone photo-fragmentation. A monoexponential fit of the three traces in Figure 4 (d) give lifetimes of 612 fs for D1, 775 fs for D2 and 1114 fs for D3 reflecting the almost exponential decay due to $S_2$ to $S_0$ internal conversion predicted by the dynamics.

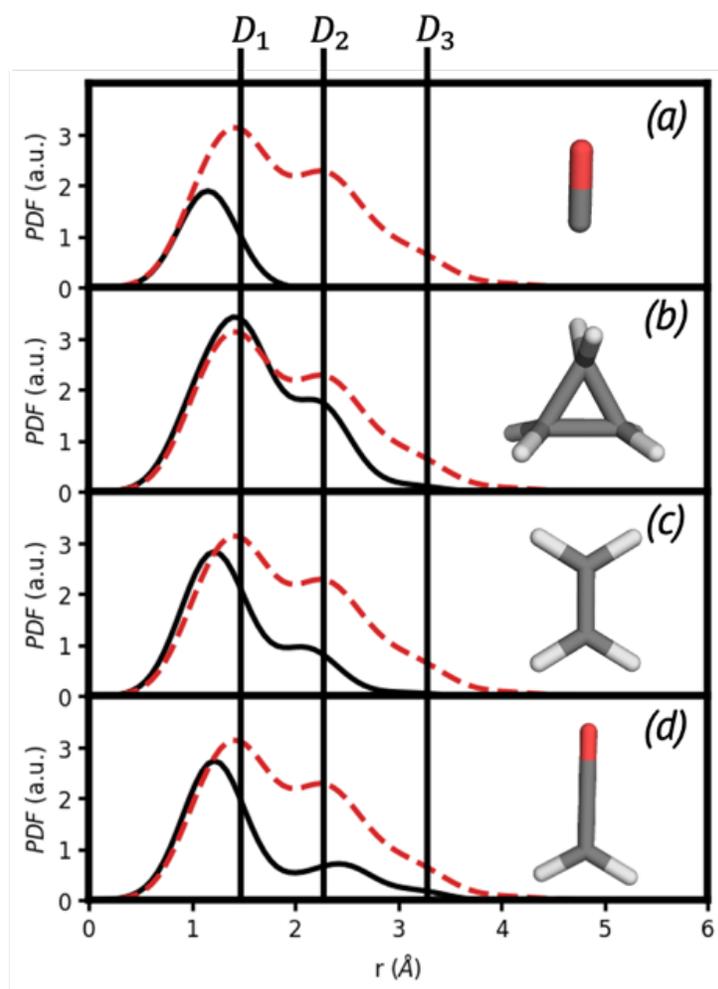

Figure 5: Individual PDF of the photoproducts formed during the ring fragmentation namely: (a) carbon monoxide, (b) cyclopropane, (c) ethylene and (d) ketene. Molecular structures of the products are reported in the inset. The $D_1$, $D_2$ and $D_3$ equilibrium distances are depicted by vertical black lines while the reference signal of the ground state is shown in each panel with dashed red lines.

At early times the loss of intensity in D1 encodes the passage from Franck Condon to the three-state degeneracy region after passage through the TS. At long times, the signal encodes information about the photoproducts. It displays a more pronounced loss of $D_2$ and $D_3$ pair distances (about 53% and 60% loss) compared to $D_1$ (only 30%). This can be explained by the expected molecular size reduction upon photo-fragmentation and quantified by comparing the PDF of the reagent with those of the photoproducts displayed in Figure 5. All the fragments show quite similar behaviors having more intense signals close to α pair distances but less intense features in $D_2$ and almost negligible in $D_3$ region, leading to the observed distribution pair distances loss. Moreover, ethylene and ketene are characterized by double C-C bonds which leads to shorter $D_1$ pair distances compared with cyclobutanone (as depicted in Figure 5 (c) and (d)) thus contributing to the gain observed in the short-range distances (the region around 0.6 Å). Similar consideration also applies to the C-O distance in carbon monoxide which is slightly shorter with respect to that one observed in the reagent.

The gain observed at large r values (greater than $D_3$) can be attributed to the pair distances that are elongating during the ring fragmentation. However, the gain of such pair distances can be recorded only until they remain in the experimentally probed window (i.e., within the electron beam coherent length).

The weak intensity of this gain signals reflects the almost incoherent nature of the $S_2$ decay, dictated by the abovementioned barrier between the $S_2$ state and the $S_2/S_1/S_0$ degeneracy region, that traps the wave packet in the $S_2$ minimum leading to a large variance of times required to reach the degeneracy region (and thus making the gain signal broad in time). However, upon accessing this part of the energy landscape, the formation of photoproducts occurs quickly, causing the breaking bond pair distances to rapidly move out of the probed window and their corresponding signals to be washed out from the diffraction pattern (decreasing the intensity of the gain signal).

## IV. CONCLUSIONS

We simulated the ultrafast photochemistry of cyclobutanone excited to n→(3s) Rydberg ($S_2$) state through mixed quantum-classical dynamics using a surface-hopping scheme using perturbatively corrected multireference wavefunctions for the electronic states involved in photorelaxation and modeled the resulting ultrafast time-resolved electron-diffraction spectra. A single exponential fit of the population of the $S_2$ state gives a lifetime of ~822 fs (~1089 fs for a red shifted pump). The major photoproducts formed are cyclopropane+CO (C3 species) and ethene+ketene (C2 species) with 40% yield each, along with ~5% population of propylene and reformed cyclobutanone. We observe no significant population of the triplet states in our simulations.

The UED pattern has been simulated using the independent atom model starting from the nuclear configurations obtained in the surface-hopping dynamics. The predicted UED signal displays a strong loss of cyclobutanone equilibrium distances which can be experimentally used to track the occurrence of photo-fragmentation and eventually to extract the time scale of the $S_2/S_0$ internal conversion.

The predominant pathway for photoproduct formation involves arrival at a 3-state degeneracy region after overcoming a transition-state lying 0.19 eV above a local minimum for the n→(3s) Rydberg state on $S_2$ adiabatic surface. This passage is enabled by the elongation of an α C-C bond combined with the out of plane motion of -CH2- moiety involved in bond elongation. These structural changes drastically change the diabatic nature of the lowest adiabatic surfaces and enable the population of biradical species which lead to photoproduct formation. We document two biradical species named as biradical-**1** biradical-**2** in this work, the former of which leads to the formation of both C2 (ethene + ketene) and C3 (cyclopropane + CO) photoproducts, while the latter exclusively leads to formation of C2. The minor reaction pathways involves traversing the conical intersection between the n→(3s) Rydberg and n→π* state lying 0.44 eV above the the n→(3s) Rydberg local minima on the $S_2$ surface.

## TABLES

Table I Vertical Excitation energies at Franck-Condon (eV)

| State | CASSCF | XMSPT2 | CASSCF | XMSPT2 |
|---|---|---|---|---|
| | |12,12| | |8,8| | |
| $S_1$ n→π* | 4.41 | 4.22 | 4.24 | 4.19 |
| $S_2$ n→Ryd | 6.52 | 6.30 | 6.21 | 6.45 |
| $T_1$ n→π* | 4.02 | 4.02 | 4.03 | 4.04 |
| $T_2$ π→π* | 5.70 | 6.21 | 5.69 | 6.26 |
| $T_3$ n→Ryd | 6.42 | 6.27 | 6.45 | 6.27 |

Table II Distribution of ground-state photoproducts at end of 2ps. ~20 percent of trajectories remain trapped in excited state at end of 2ps.

| Product | % |
|---|---|
| CO+cyclopropane | 40 |
| Ethene + ketene | 38 |
| CO+propylene | 5 |
| cyclobutanone | 4 |

Table IIII Value of SOC[a] at critical structures and including vibronic effects (cm-1)

| States | FC | S2-min | S2-min + vibronic |
|---|---|---|---|
| S2-T1 ($^3$nπ*) | 1.546 | 1.624 | 2.72 |
| S2-T2 ($^3$ππ*) | 1.762 | 0 | 0.78 |

[a] The value of SOC is the sum over the SOC with three states of the triplet

## SUPPLEMENTARY MATERIAL

In the Supplementary Material we provide:

Active-space orbitals utilized in in |12,12| and |8,8| active-spaces, population dynamics for mixed quantum-classical dynamics at |12,12| CASSCF level, optimized IRC at |12,12| CASSCF level, structure of the optimized TS on S2 surface and the negative normal mode, the optimized reaction path at XMS-CASPT2 level for the minor photorelaxation pathway, the reaction path at XMS-CASPT2 level in case of triplet population, prediction of scattering intensity at FC and along photorelaxation, UED signals with infinite time-resolution, UED signals from CASSCF dynamics, cartesian coordinates for critical structures, plots of energies of electronic states in trajectories that hop to $S_0$ around the time of hopping.

## ACKNOWLEDGEMENTS

We acknowledge support from the U.S. Department of Energy, Office of Science, Office of Basic Energy Sciences, Chemical Sciences, Geosciences and Biosciences Division under award no. DE-SC0022225.

## DATA AVAILABILITY STATEMENT

The data that support the findings of this study are available within the article and its supplementary material.

## PRE-PRINT SUBMISSION

A copy of this manuscript is available on arXiv: http://arxiv.org/abs/2402.09873

# SUPPLEMENTARY

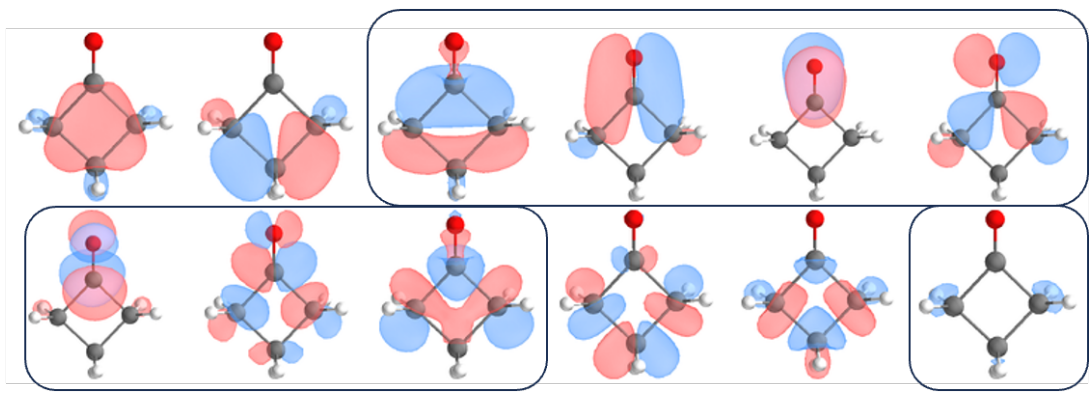

Figure S1: CASSCF |12,12| and CASSCF |8,8| active-spaces utilized in this work

# XMS-CASPT2 |12,12|

|  | 6-31G*/8s8p8d | 6-31G*/1s1p1d | 6-31++G* | Aug-cc-pvdz | Aug-cc-pvtz |
|---|---|---|---|---|---|
| $^1$NPI* | 4.24 | 4.34 | 4.32 | 4.25 | 4.24 |
| $^1$N RYDBERG | 6.28 | 6.36 | 6.34 | 6.24 | 6.50 |
| $^3$NPI* | 4.03 | 4.02 | 4.02 | 3.93 | 3.92 |
| $^3$PIPI* | 6.28 | 6.25 | 6.21 | 6.21 | 6.18 |
| $^3$N RYDBERG | 6.16 | 6.24 | 6.27 | 6.10 | 6.36 |

Vertical excitation energies in eV on MP2 OPTIMIZED geometry at XMS-CASPT2 level with |12,12| active-space with various basis-sets employing diffuse functions.

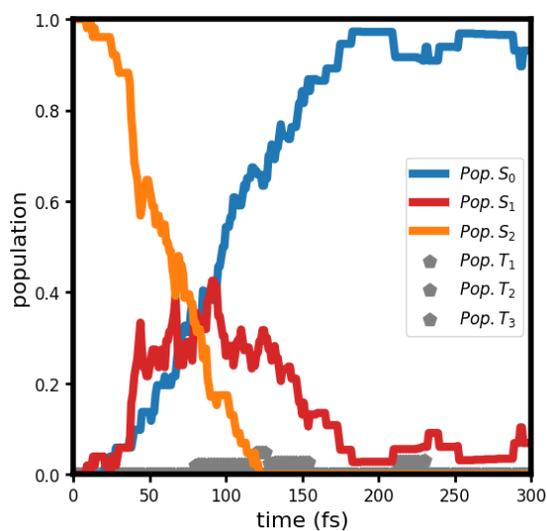

Figure S2: CASSCF |12,12| dynamics. Time dependent population of the three lowest singlet and triplet adiabatic states computed through average occupation numbers in the ensemble of trajectories.

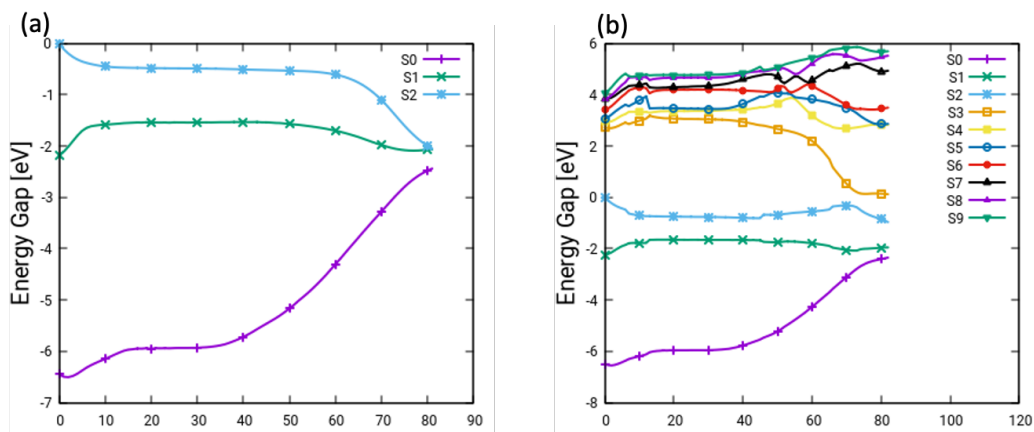

Figure S3: (a) CASSCF |12,12| IRC path. (b) Single points with 10 electronic states computed at |12,12| CASSCF level on CASSCF IRC showing the influence of higher electronic states at Franck-Condon in the IRC

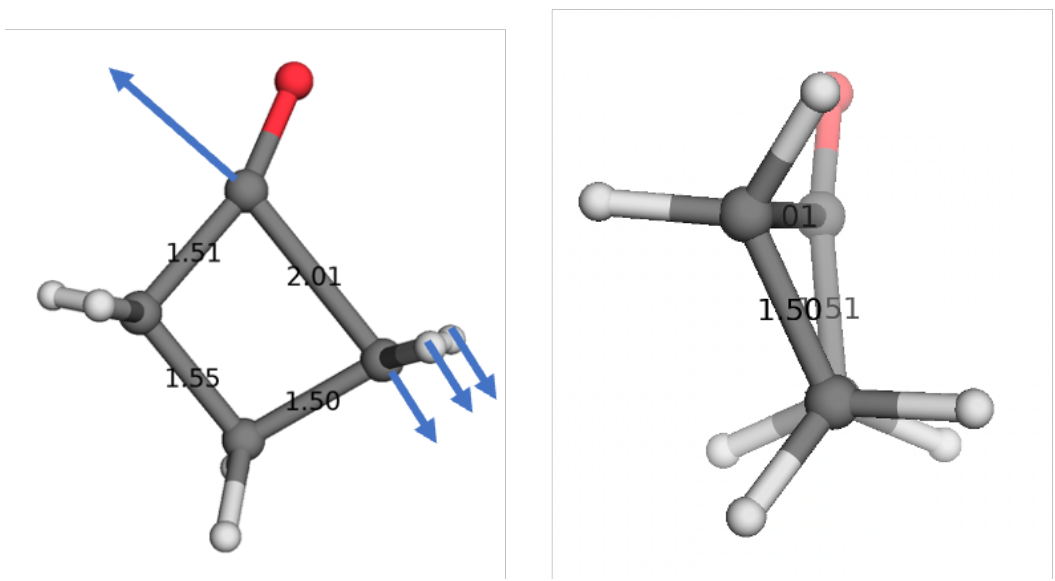

Figure S4: Normal mode corresponding to the negative frequency at 608 cm$^{-1}$ for the Transition-State on the S2 surface optimized at XMS-CASPT2 level.

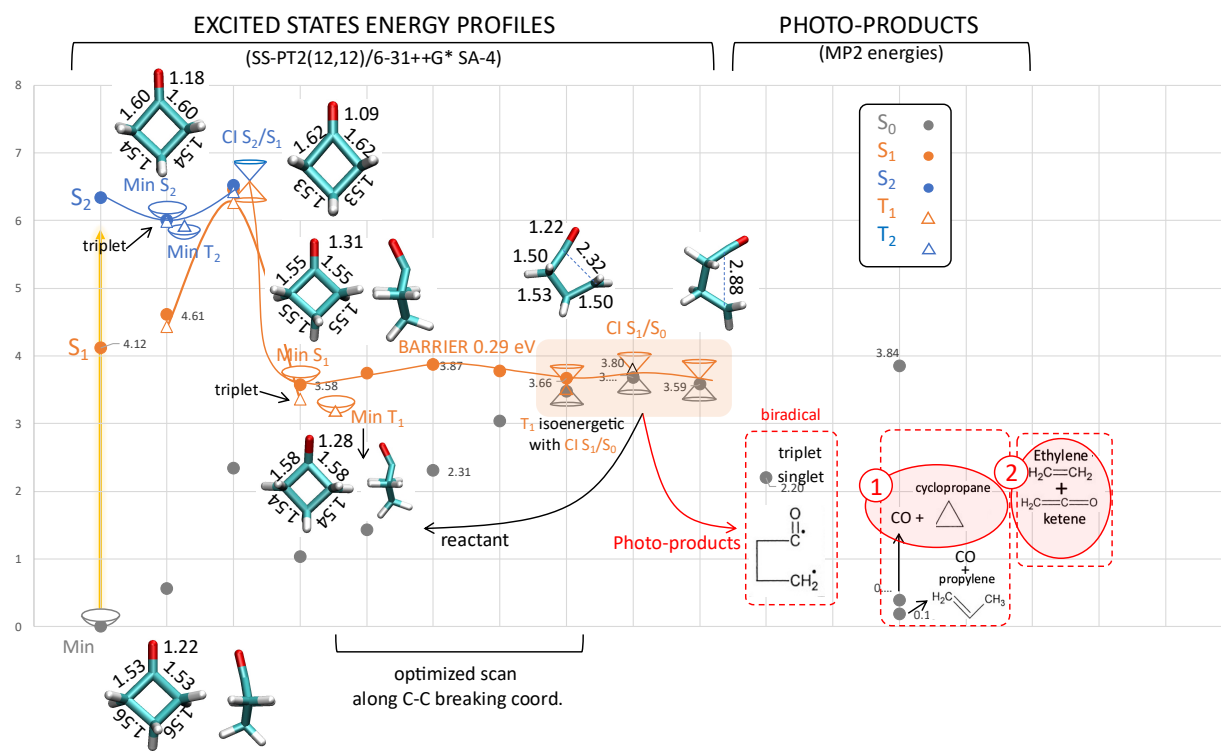

Figure S5: The minor reaction path corresponding to the CI between the nπ* and n→(3s)Rydberg state

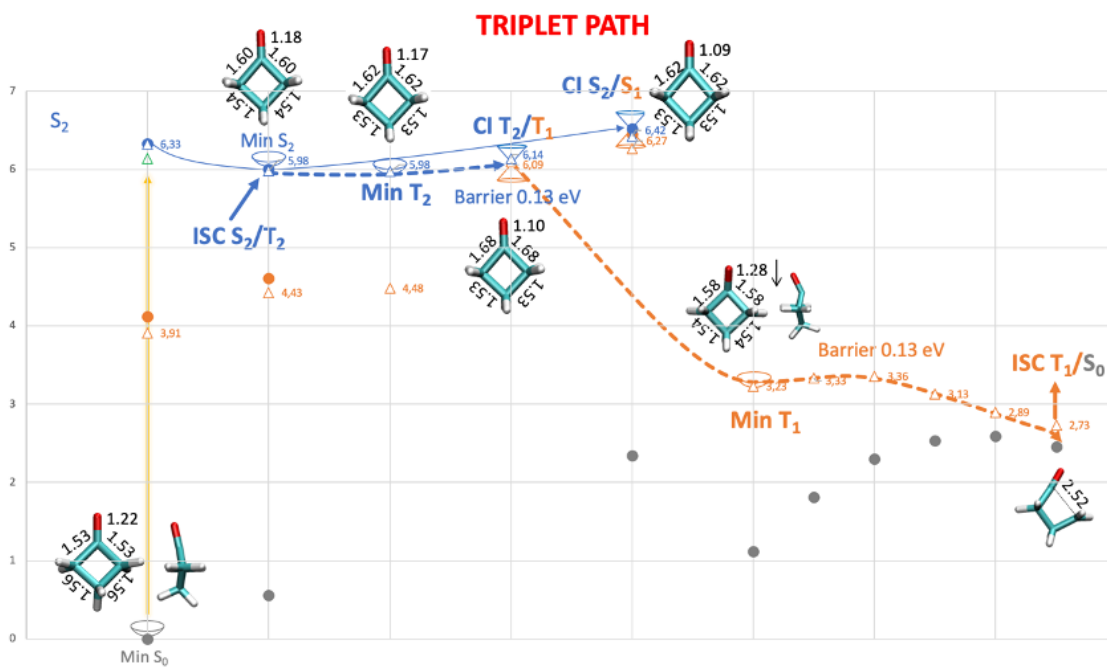

Figure S6: The reaction path in case of population of the T2 state while the dynamics are trapped in the region of S2-minima.

**XMS-CASPT2 modified scattering intensity**

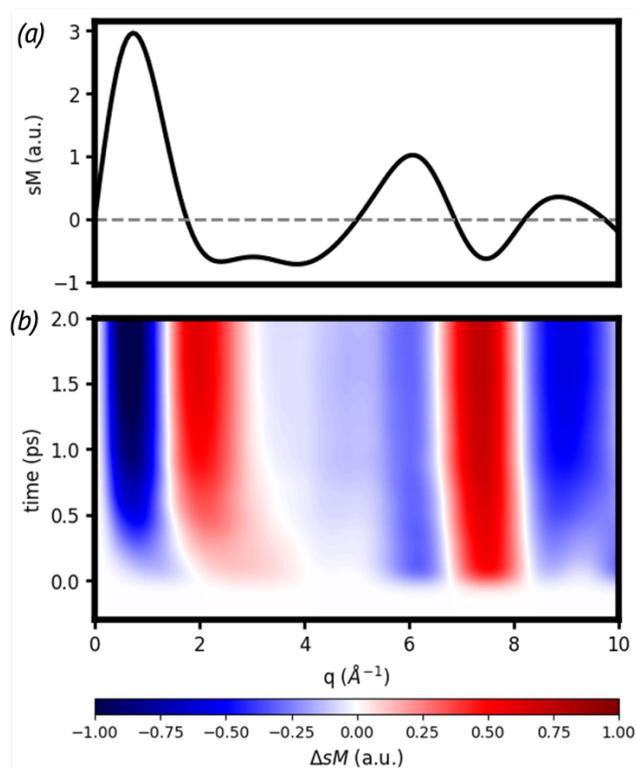

Figure S7 (a) Ground state modified scattering intensity (sM(q)) and its time dependent version (b) computed starting from XMS-CASPT2 trajectories. The predicted results have been convoluted with a gaussian function with FWHM of 150 fs mimicking the experimental time resolution.

## XMS-CASPT2 UED signals with infinite time resolution

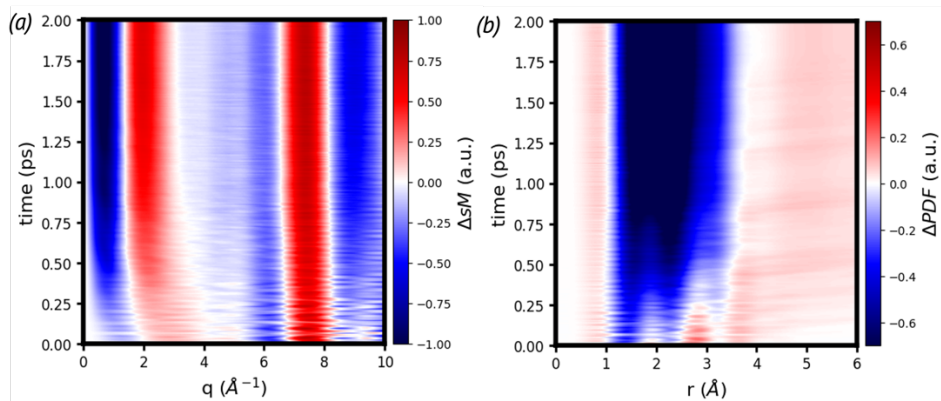

Figure S8. (a) predicted modified scattering intensity $\Delta sM(q)$ and (b) time dependent pair distribution function $\Delta PDF(r)$ computed from XMS-CASPT2 trajectories without accounting for the experimental time resolution.

## UED signals from CASSCF trajectories

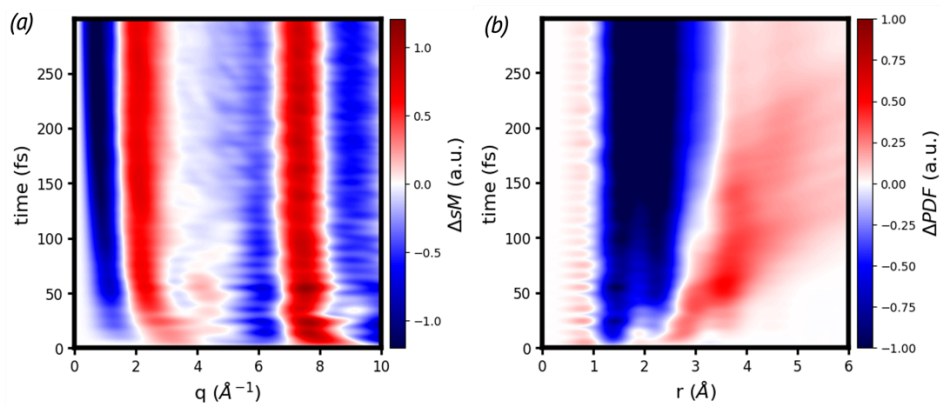

Figure S9. (a) predicted modified scattering intensity and (b) time dependent pair distribution function computed from CASSCF trajectories without accounting for the experimental time resolution.

**Comparison between IAM and ab-initio UED simulations**

In this section we simulate UED signals from ab-initio electron densities computed for three different CASSCF(12,12) IRC steps (namely: 15, 45 and 80) by post processing them as described in [doi.org/10.1063/4.0000043]. These scattering patterns are compared with the signals predicted by the IAM to access the strength of the individual atom approximation.

Fig. S10 shows that the main differences between IAM and ab-initio signals are observed for momentum transfer values (q) smaller than 5 Angstrom while, for large q values, the two approaches look almost identical. This behavior is expected since at large momentum transfers the main contribution to the UED signal comes from the nuclear scattering (i.e., those electrons that are scattered directly by the nuclei) which is properly captured already in the IAM. The difference between the IAM and ab-initio approach at low momentum is thus originated by the electron-electron and mixed electron-nuclear scattering interactions which are better described when ab-inito electron densities are employed. However, these differences become even less relevant when looking to the Fourier transform of the modified diffraction patterns (i.e., the pair distribution function in the real space) since they do not affect the global behavior of the signal.

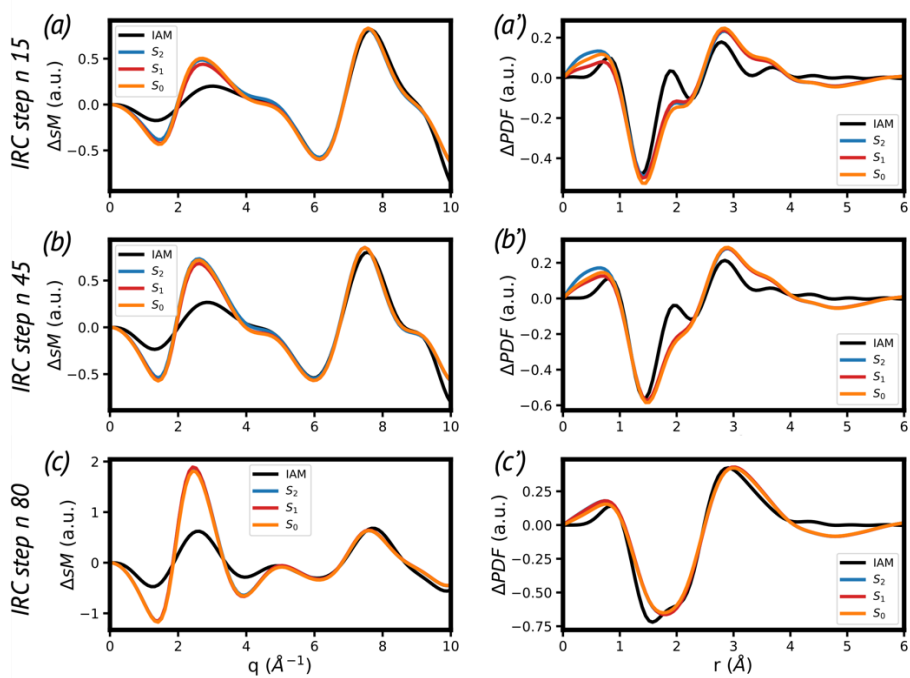

Figure S10. Comparison between the IAM and ab-initio UED simulations performed on some structures of the CASSCF(12,12) IRC path (i.e., step number 15 (a/a'), 45 (b/b') and 80 (c/c') of figure S3-(a)). In particular, a,b, and c compare the modified scattering intensity predicted trough IAM with those obtained employing the electron densities of the three main electronic states involved in the ring fragmentation process ($S_2$, $S_1$ and $S_0$). Panels, a',b' and c', analogously, compare the Fourier transform of the modified scattering intensity (i.e., the radial pair distribution function) obtained with the two different approaches.

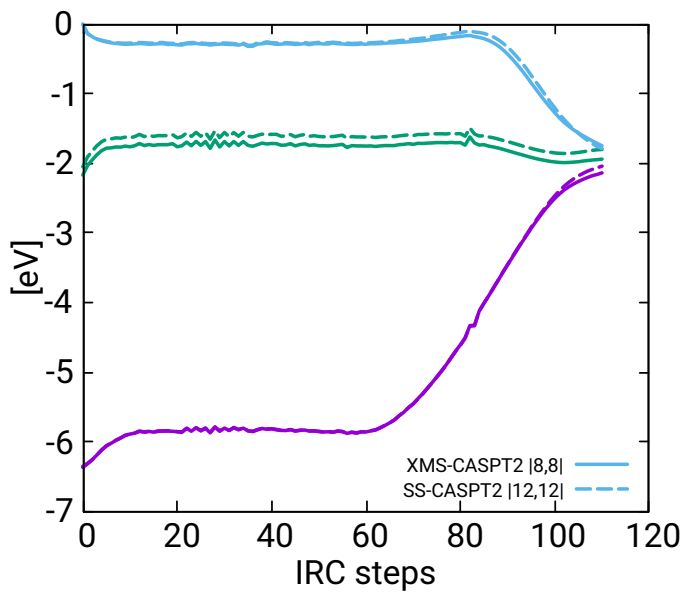

Figure S11: Comparison of the energetic profile of relevant electronic states along major reaction path with two different active-spaces at CASPT2 level.

(a)

(b)

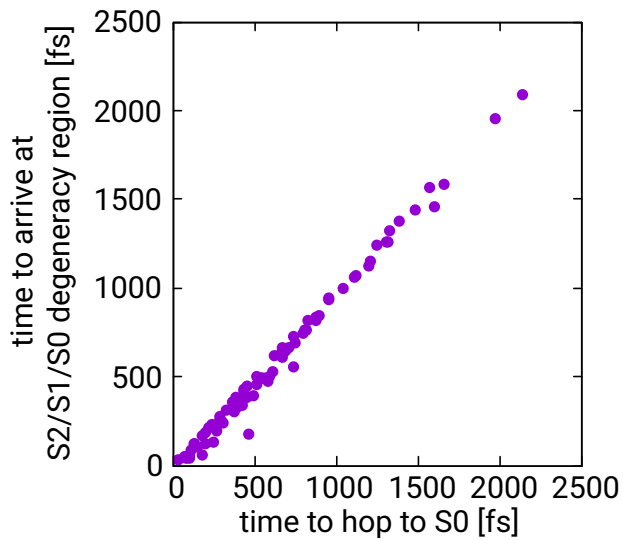
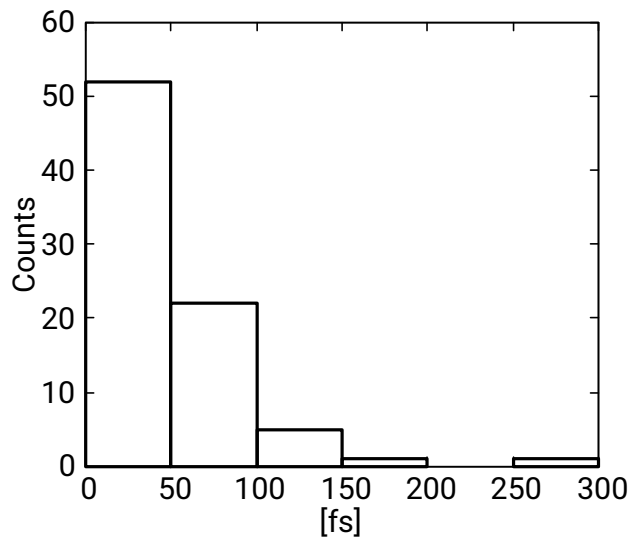

Figure S12: (a) hopping time vs the time to arrive at three-state degeneracy region. (b) Distribution of time spent in three-state degeneracy region

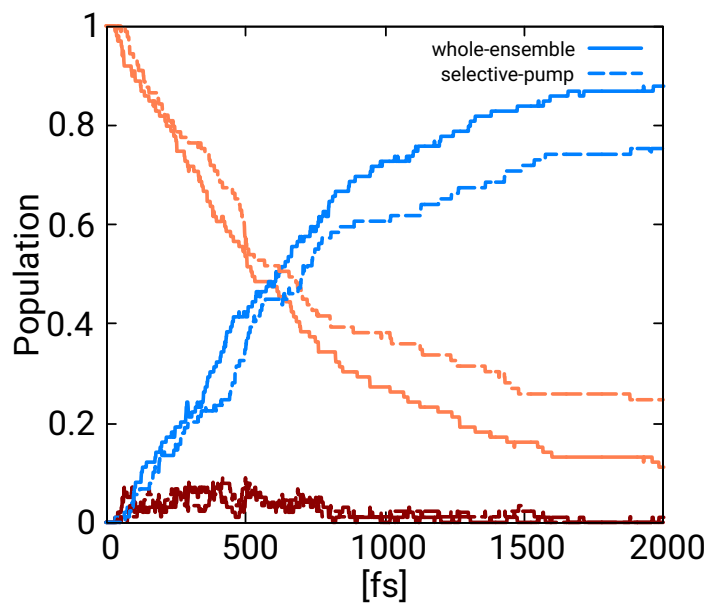

Figure S13: Comparison of non-adiabatic dynamics by selecting geometries with Rydberg excitation lying within 6-6.2 eV (*selective pump*) versus selecting all geometries in Wigner ensemble *(whole-ensemble)*. *Whole-ensemble* consists of 100 geometries while *selective-pump* ensemble consists of 89 geometries.

| Product | % |
|---|---|
| CO+cyclopropane | 41 |
| Ethene + ketene | 39 |
| CO+propylene | 4 |
| cyclobutanone | 4 |

Table S2: Quantum yield of photoproducts in selective pump dynamics

Coordinates of critical geometries optimized at CASPT2 level

```
Min S0
  C    -0.387848   -1.104413    0.074740
  C    -1.470042   -0.000011   -0.086827
  C    -0.387859    1.104436    0.074497
  C     0.658865   -0.000001   -0.091599
  O     1.849145   -0.000024   -0.351073
  H    -0.350727    1.914687   -0.659990
  H    -0.353464    1.534715    1.082566
  H    -1.914412   -0.000124   -1.085465
  H    -2.268719    0.000067    0.658052
  H    -0.350707   -1.914823   -0.659570
  H    -0.353449   -1.534471    1.082903

Min S2
  C    -0.451410   -1.082527    0.004298
  C    -1.539687   -0.000004   -0.018125
  C    -0.451421    1.082535    0.004060
  C     0.732631    0.000012    0.028149
  O     1.909024    0.000020    0.052038
  H    -0.287637    1.646588   -0.926465
  H    -0.325616    1.646289    0.940656
  H    -2.132471   -0.000107   -0.932404
  H    -2.169413    0.000090    0.871111
  H    -0.287620   -1.646783   -0.926103
  H    -0.325599   -1.646075    0.941018
```

Min S1
```
  C    -0.395994   -1.090507    0.013226
  C    -1.492874   -0.000011   -0.078396
  C    -0.396004    1.090517    0.012987
  C     0.642833   -0.000032   -0.371190
  O     1.882219    0.000020    0.043657
  H    -0.448718    1.946559   -0.665191
  H    -0.217598    1.435264    1.040107
  H    -1.994444   -0.000119   -1.048358
  H    -2.242355    0.000073    0.715734
  H    -0.448699   -1.946699   -0.664764
  H    -0.217584   -1.435027    1.040422
```

CI S2S1
```
  C    -0.485643   -1.049046    0.000404
  C    -1.597783   -0.000006   -0.035524
  C    -0.485654    1.049052    0.000177
  C     0.752086    0.000010    0.005383
  O     1.839969    0.000018    0.029564
  H    -0.220497    1.555676   -0.934188
  H    -0.275344    1.514084    0.992646
  H    -2.172636   -0.000109   -0.954888
  H    -2.187906    0.000090    0.875528
  H    -0.220483   -1.555877   -0.933847
  H    -0.275327   -1.513855    0.992978
```

MinT1
```
  C    -0.401276   -1.081174    0.017429
  C    -1.498153    0.000186   -0.078168
  C    -0.402350    1.081673    0.016937
  C     0.661802   -0.000139   -0.413575
  O     1.855994   -0.001652    0.058220
  H    -0.448083    1.941449   -0.656496
  H    -0.203154    1.405244    1.046027
  H    -1.989314   -0.000171   -1.053065
  H    -2.253714    0.000225    0.709603
  H    -0.448027   -1.941311   -0.655424
  H    -0.202942   -1.404291    1.046746
```

Min T2
```
  C    -0.459400   -1.087967    0.004294
  C    -1.538945   -0.000004   -0.018246
  C    -0.459411    1.087975    0.004055
  C     0.741036    0.000012    0.028136
  O     1.913925    0.000021    0.051423
  H    -0.286478    1.650395   -0.926578
  H    -0.324650    1.649786    0.941410
  H    -2.133470   -0.000107   -0.931151
  H    -2.170730    0.000090    0.869334
  H    -0.286462   -1.650591   -0.926215
  H    -0.324633   -1.649572    0.941773
```

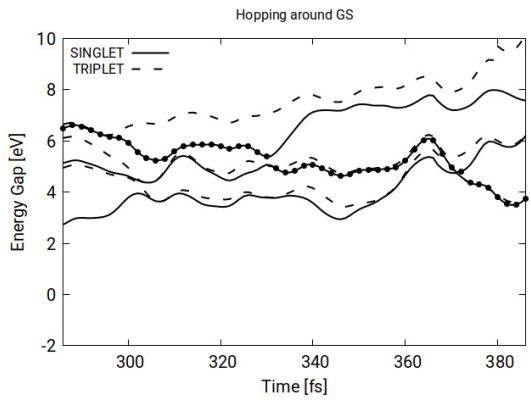
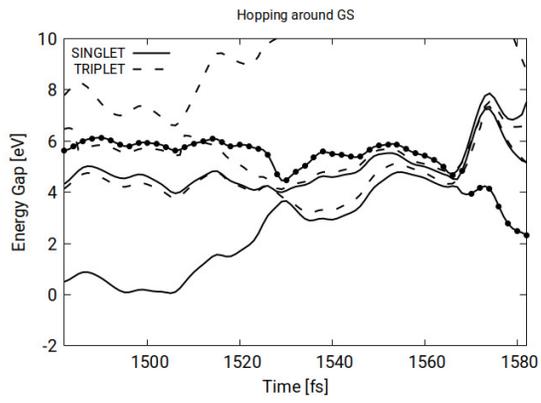
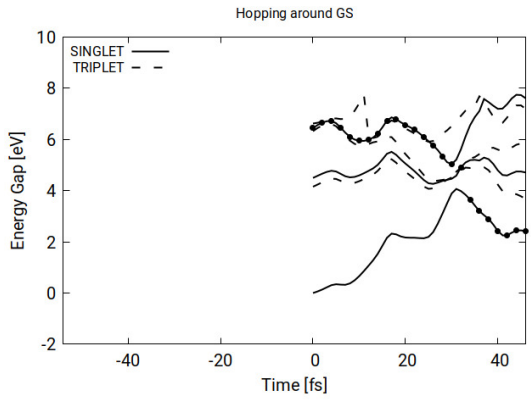
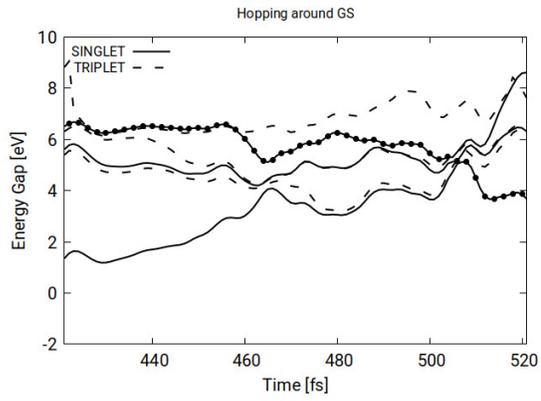
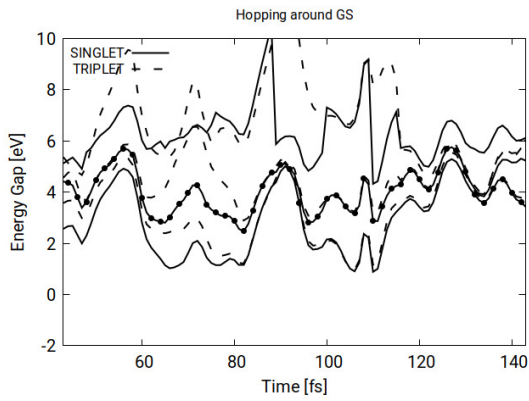
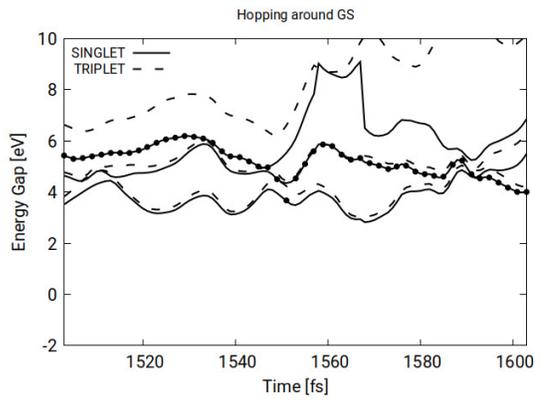
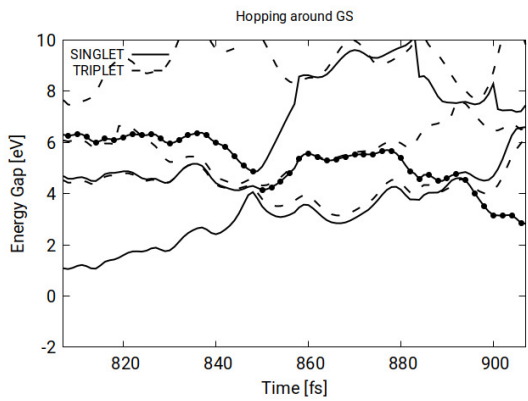
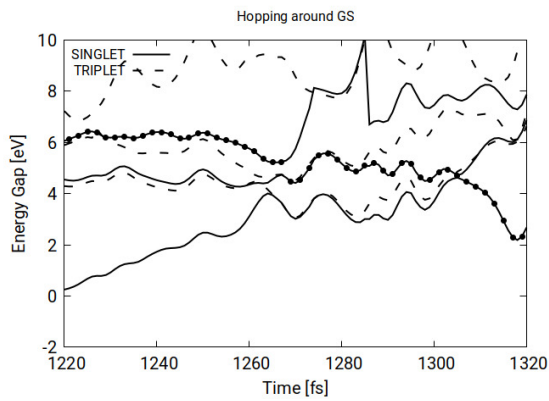

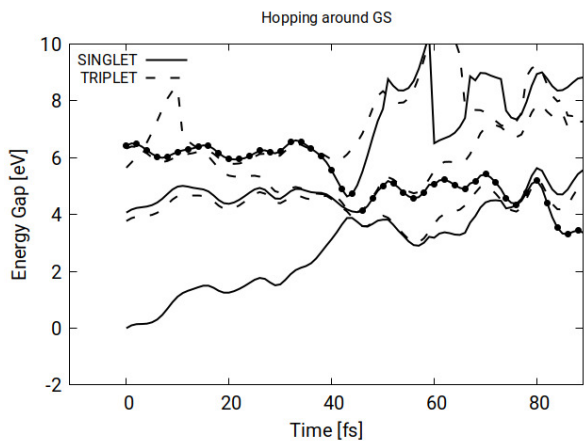
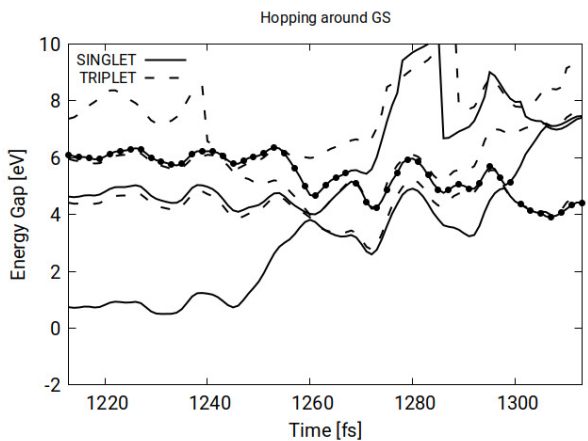
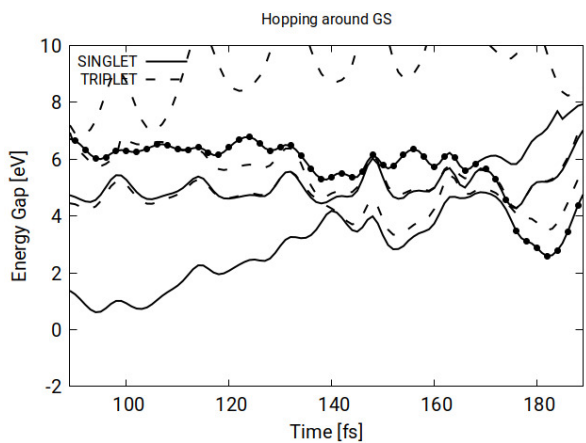
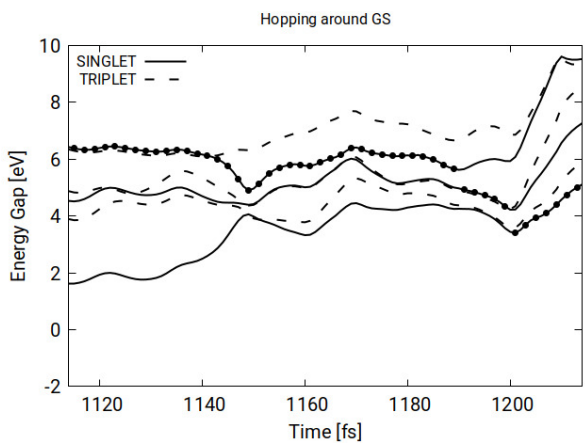
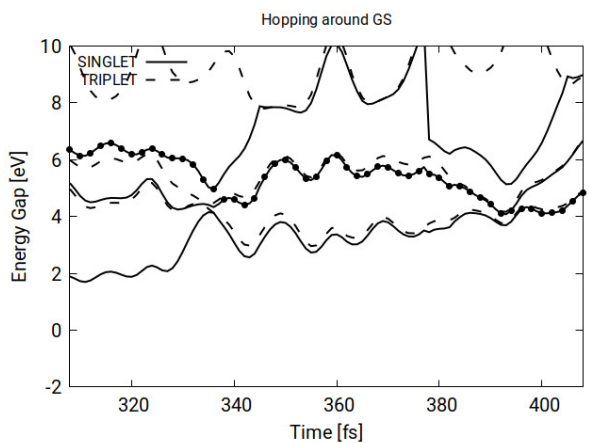
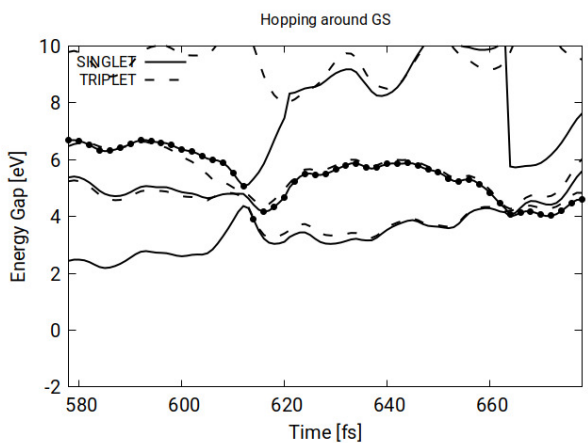
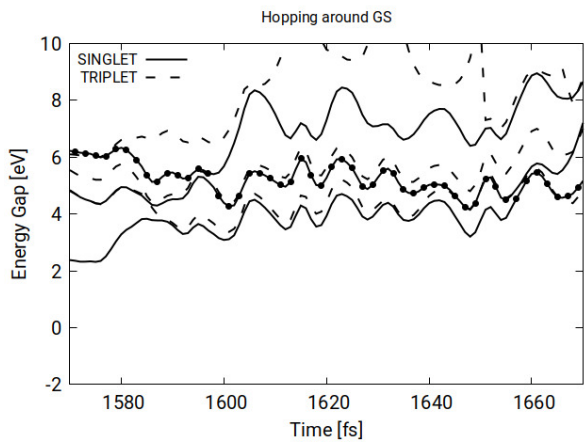
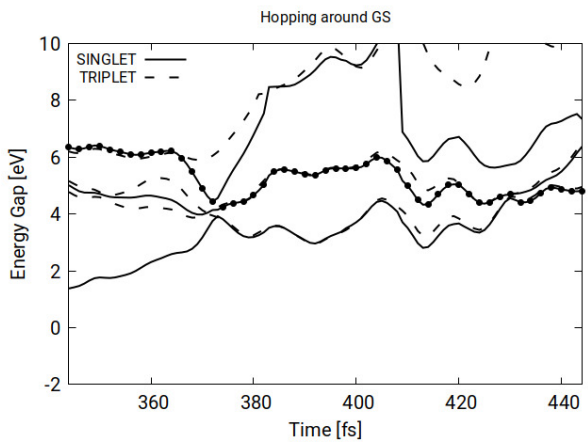

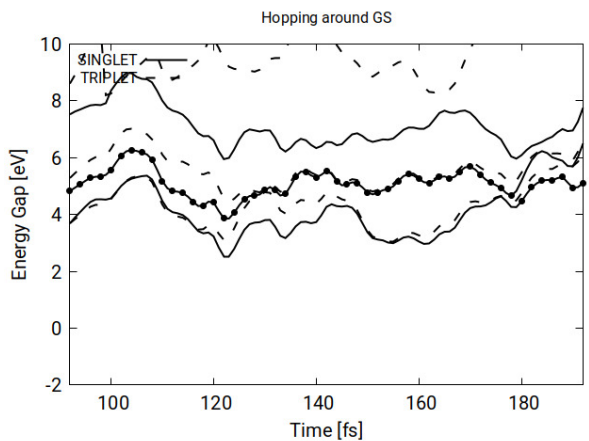
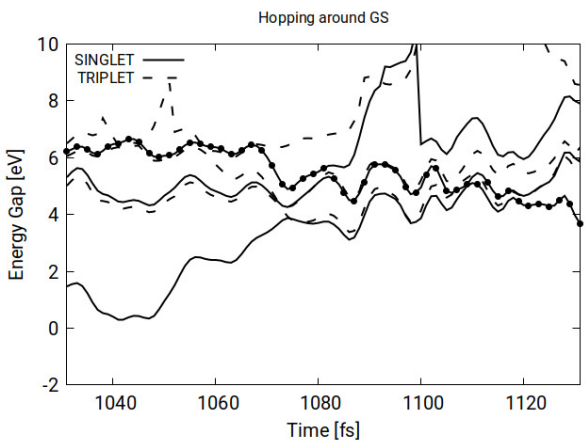
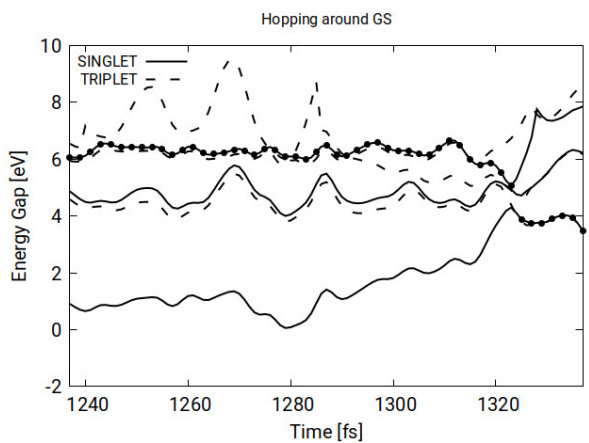
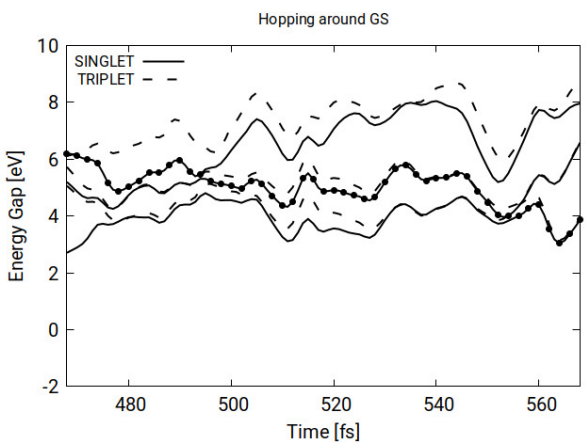
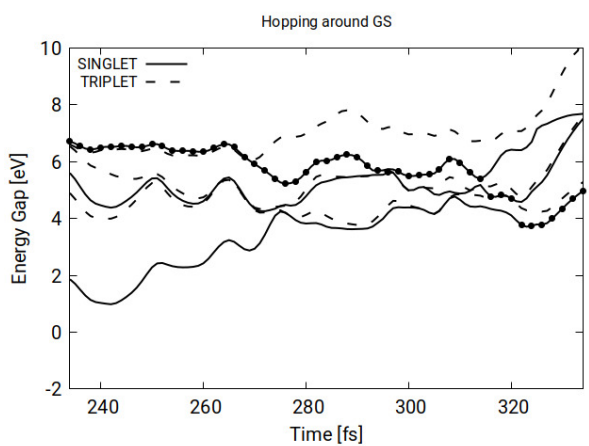
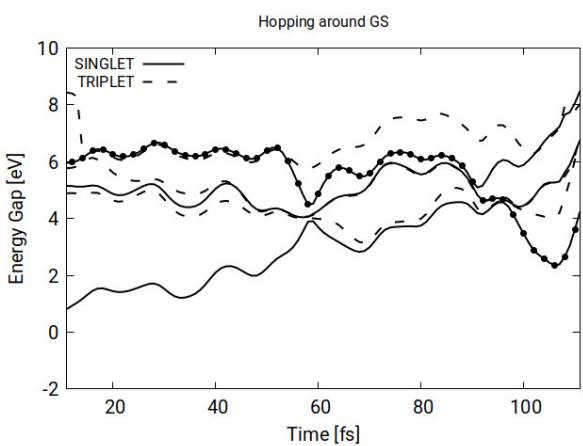
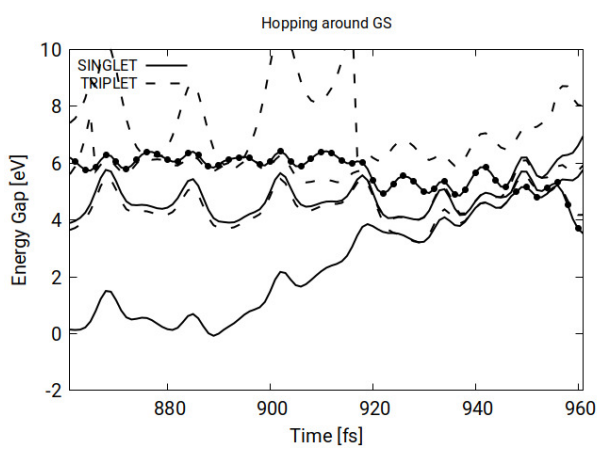
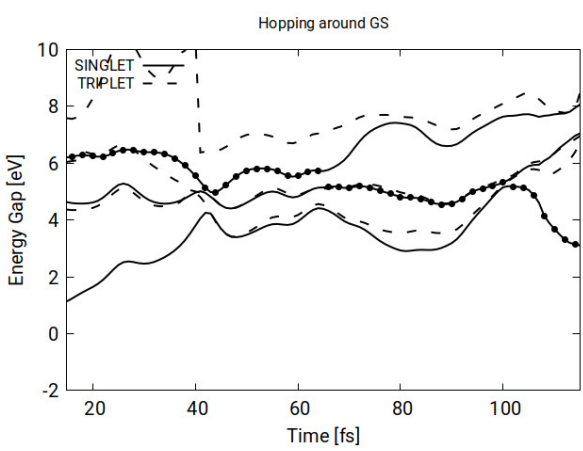

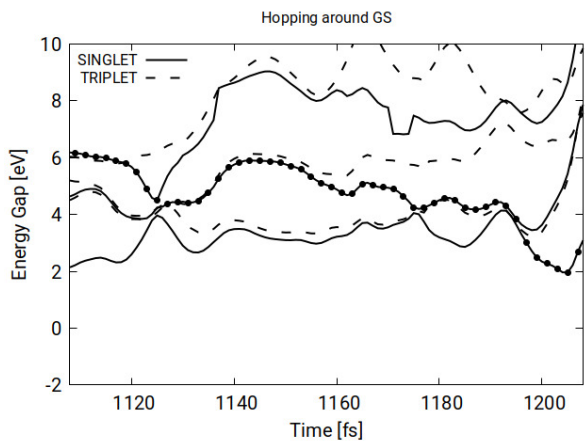
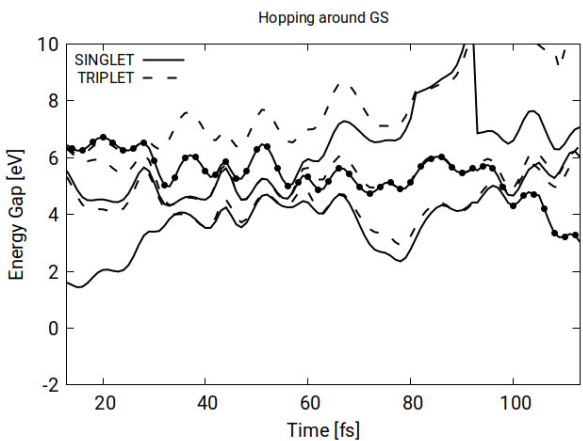
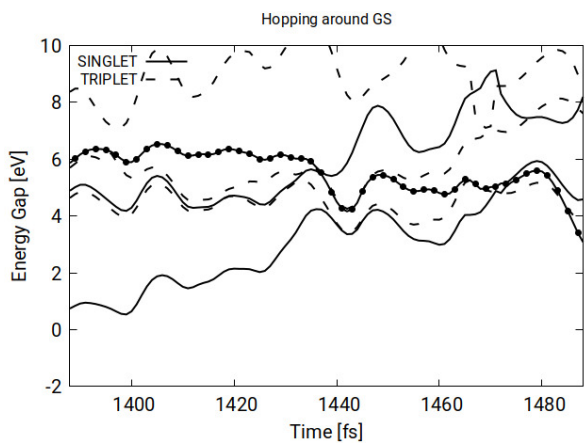
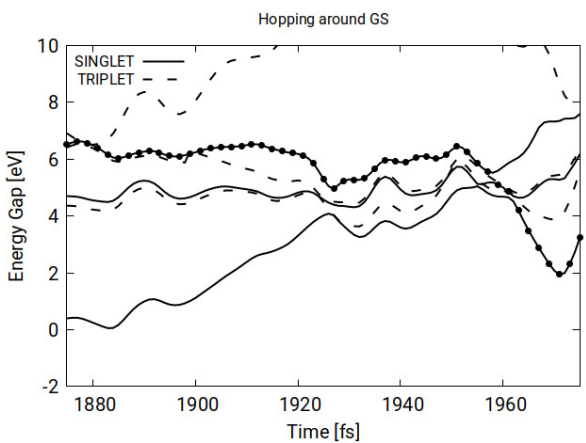
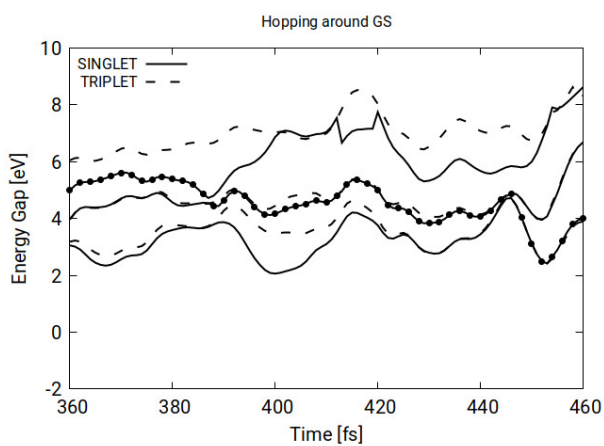
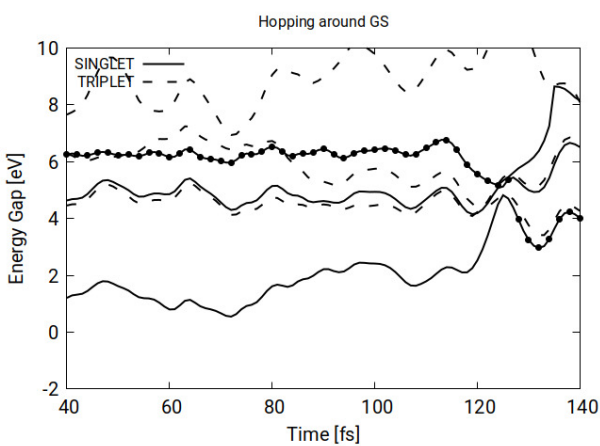
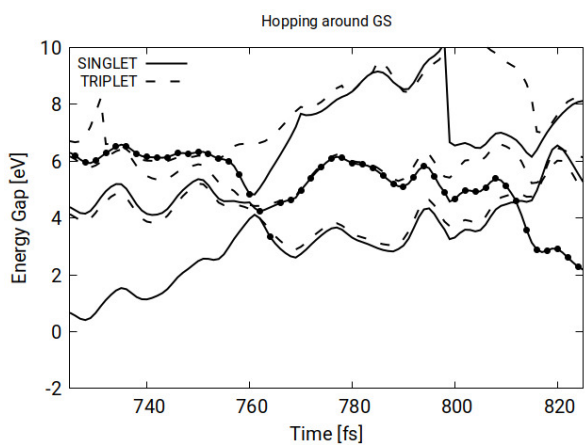
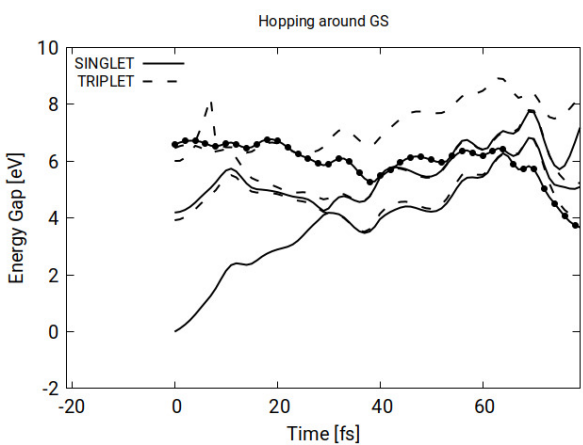

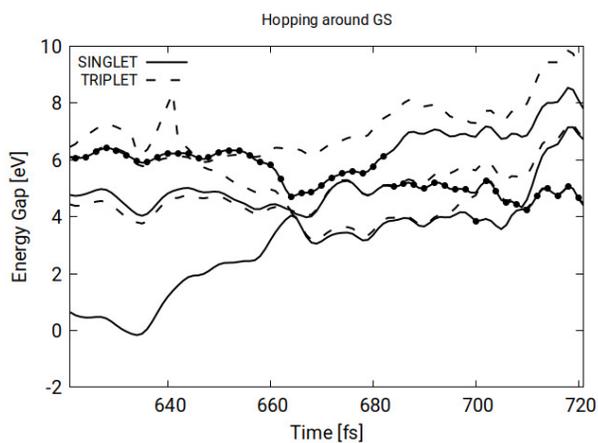
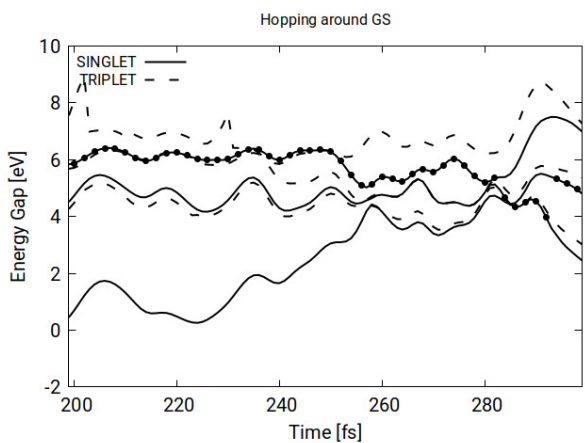
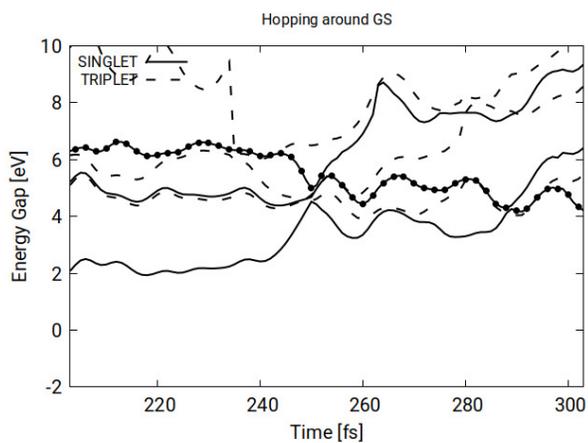
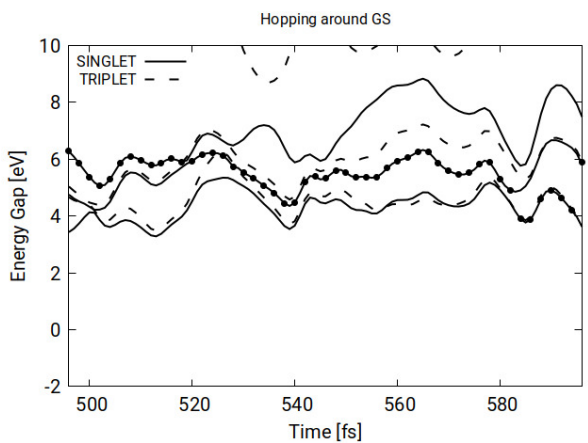
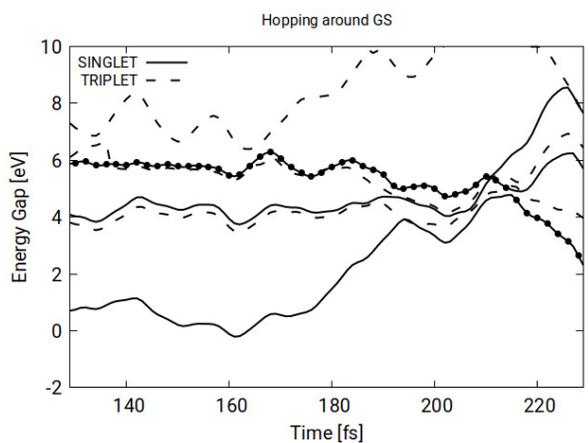
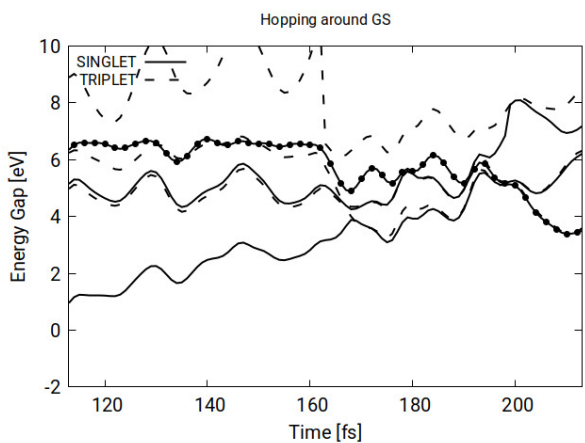
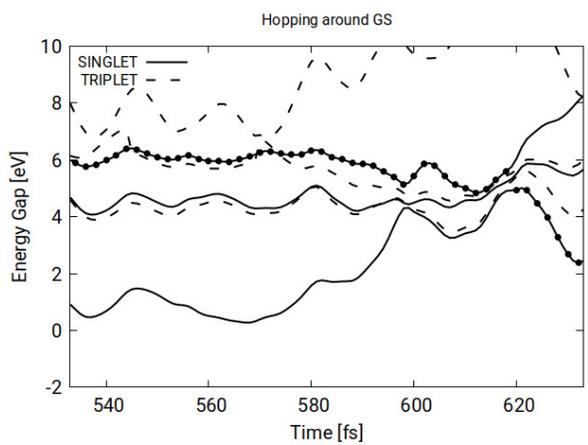
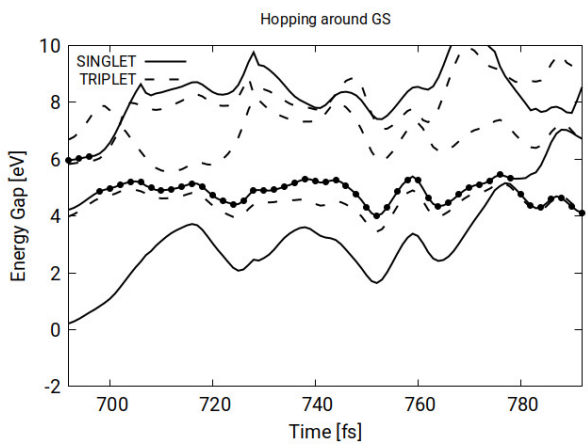

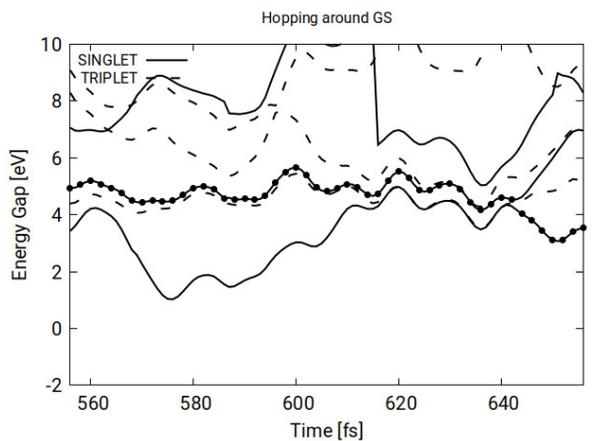
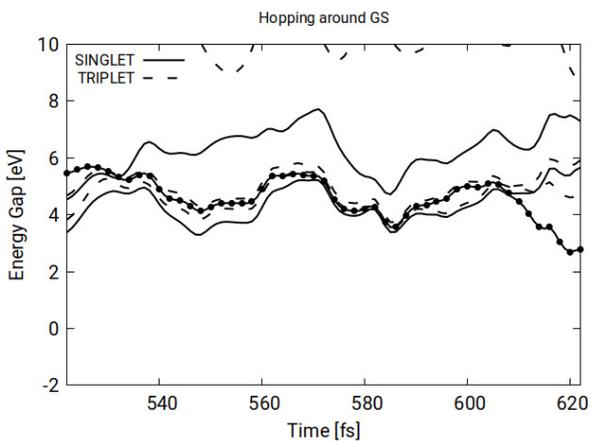
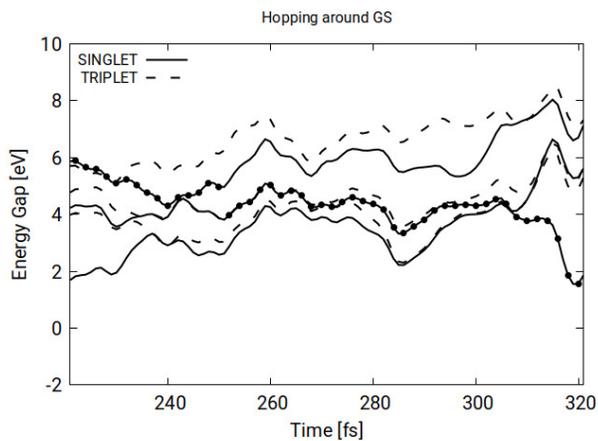
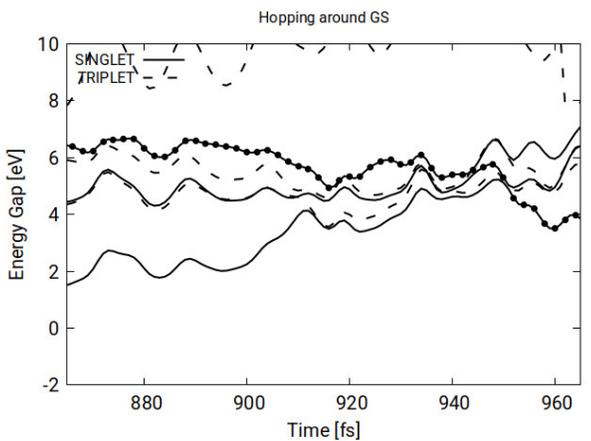
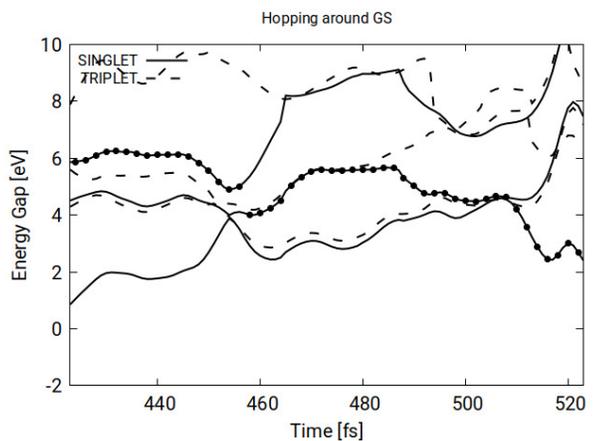
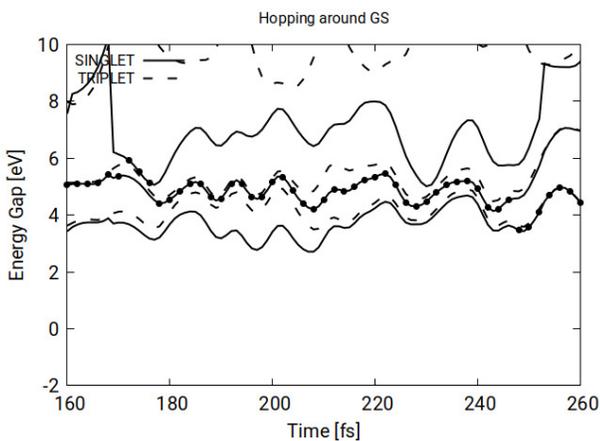
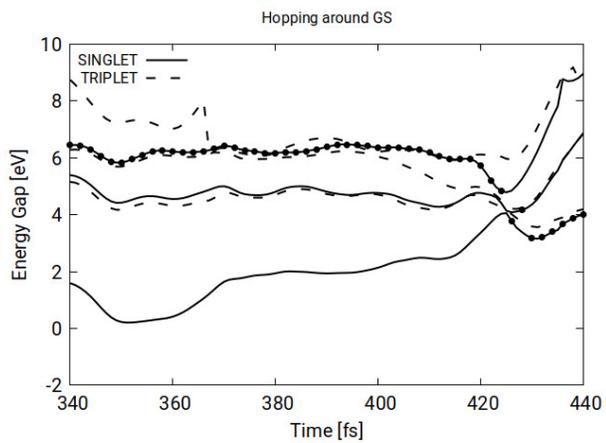
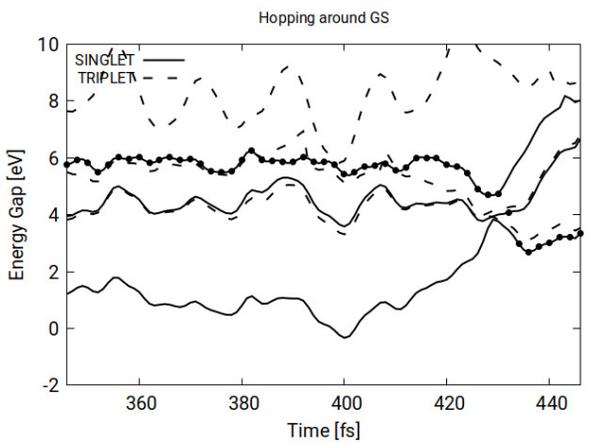

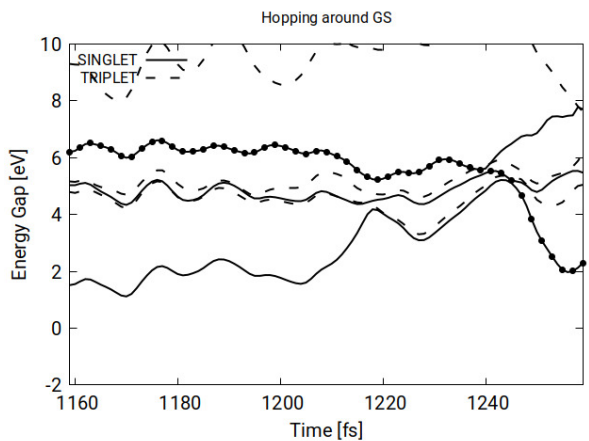
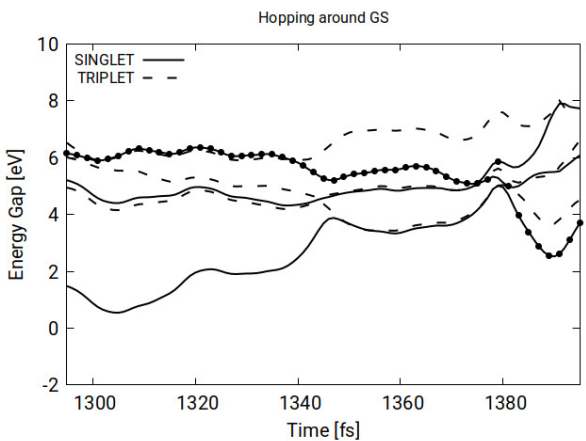
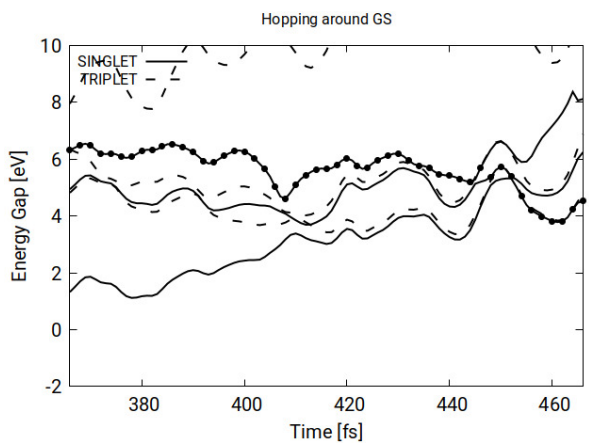
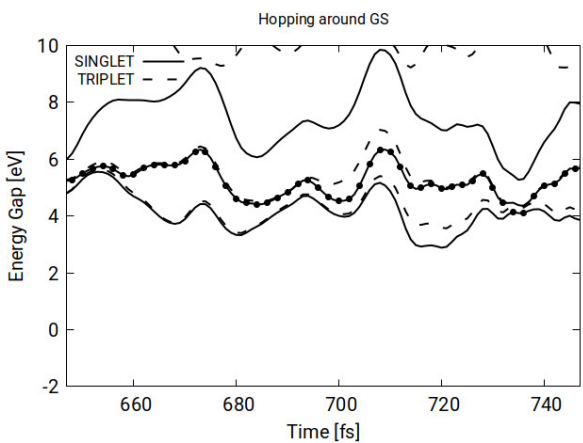
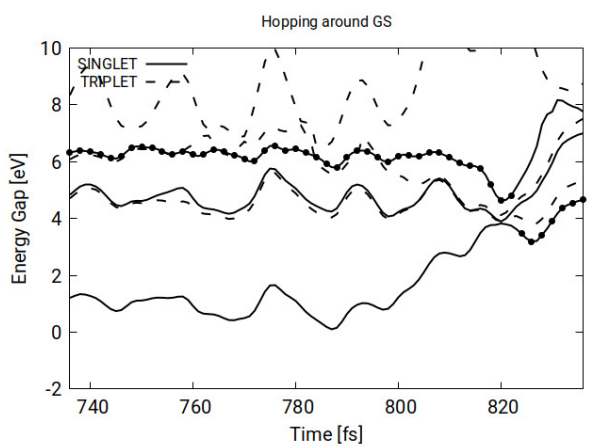
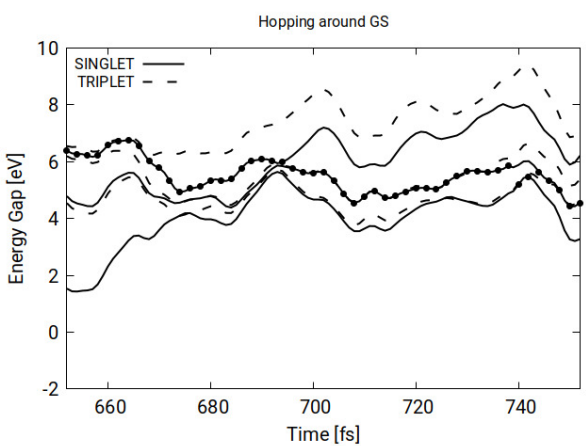
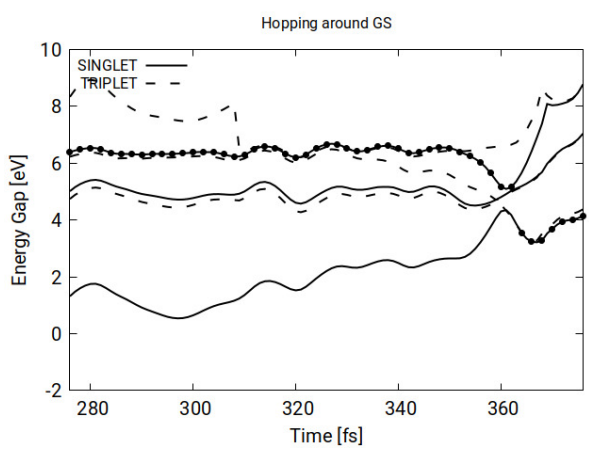
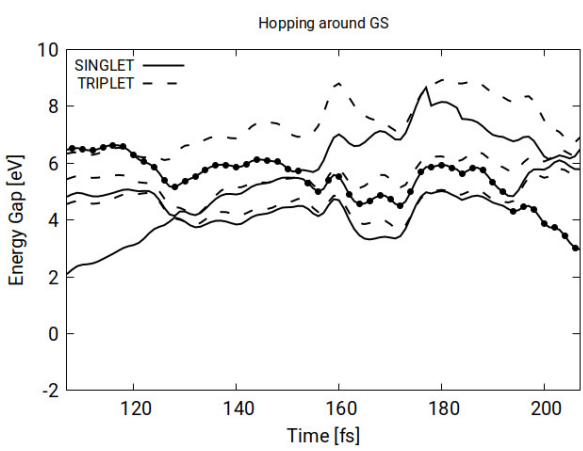

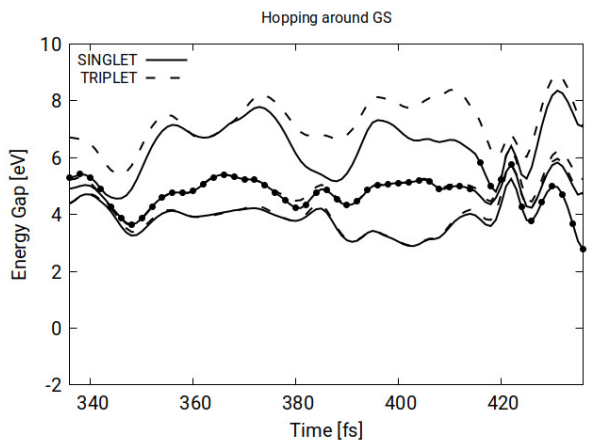
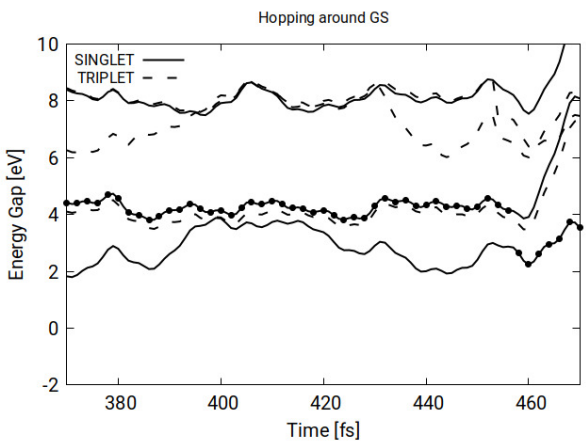
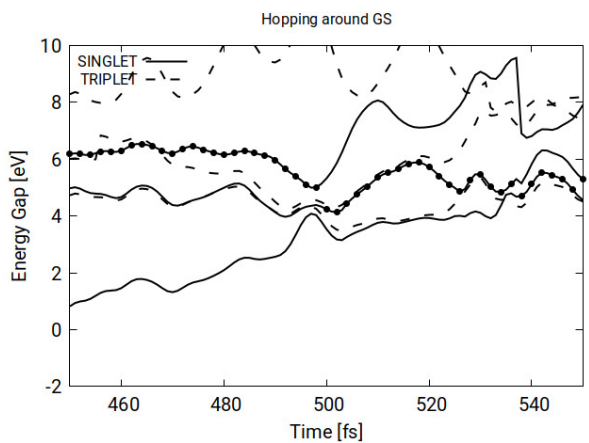
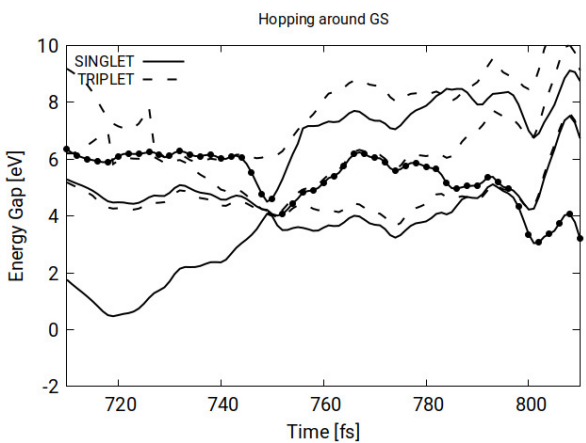
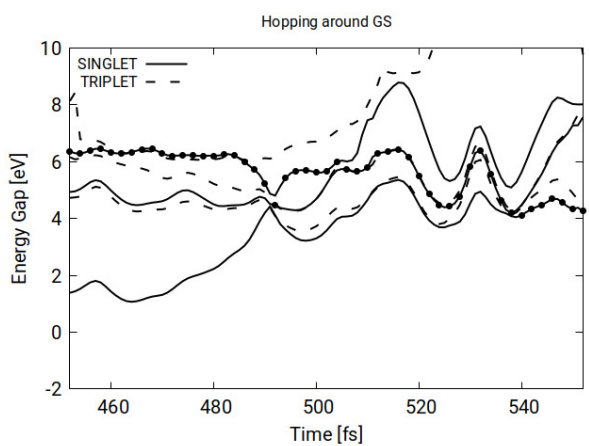
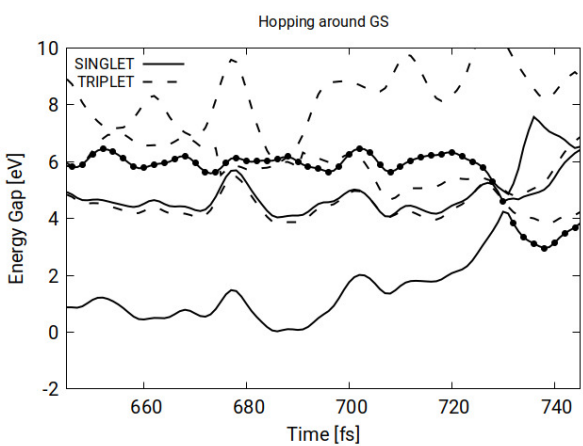
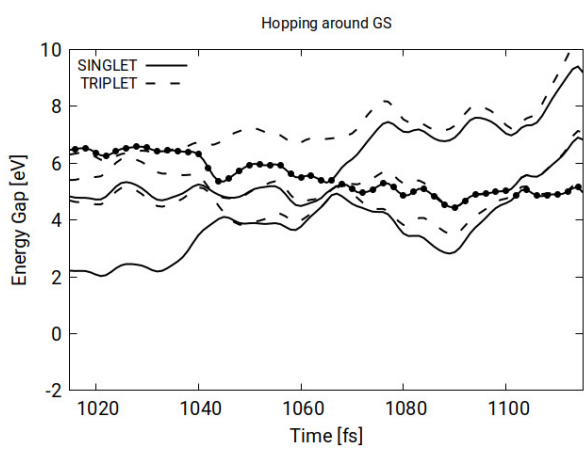
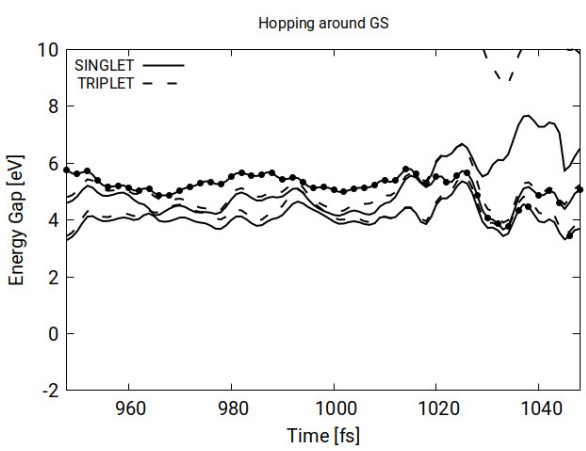

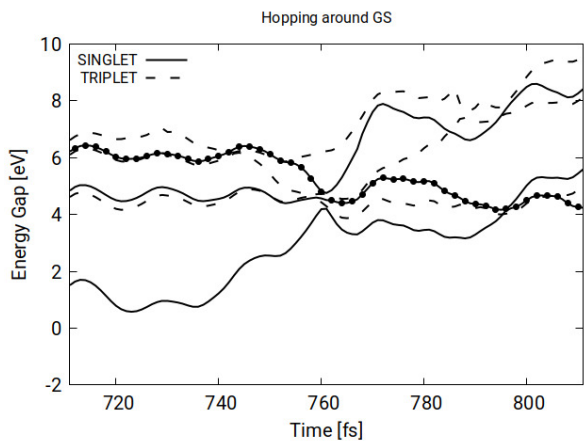
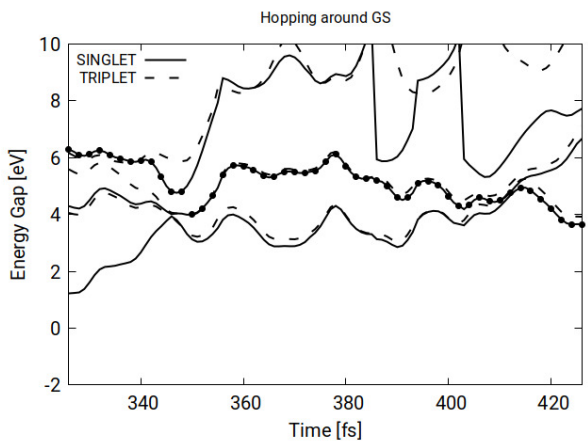
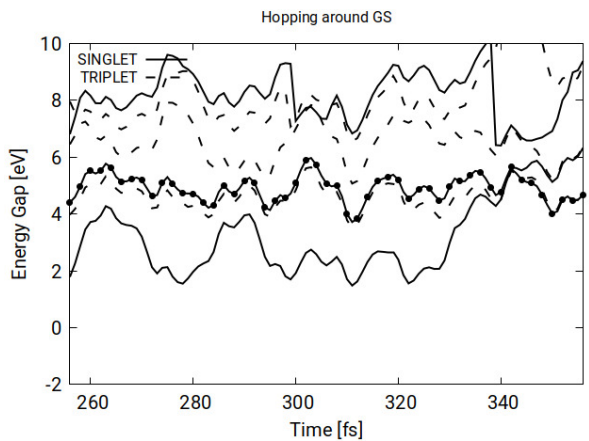
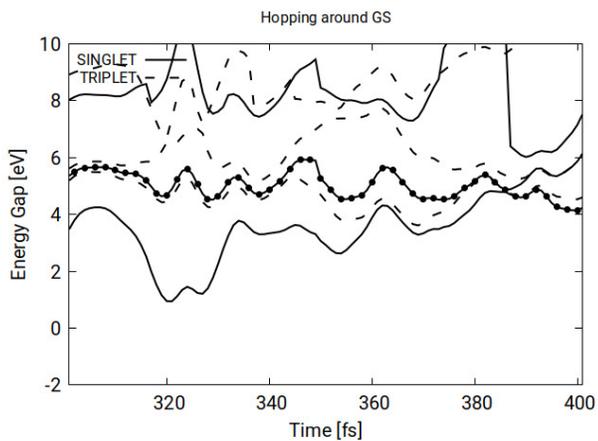
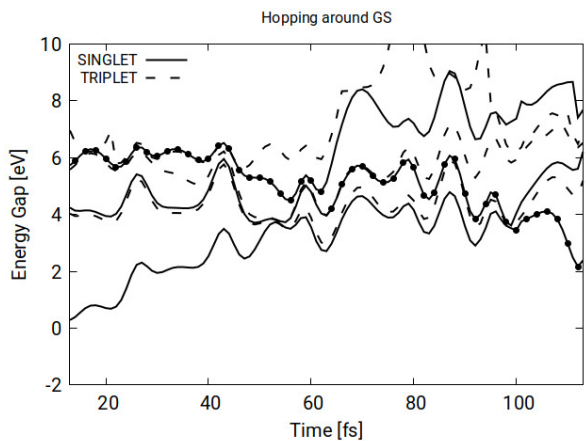
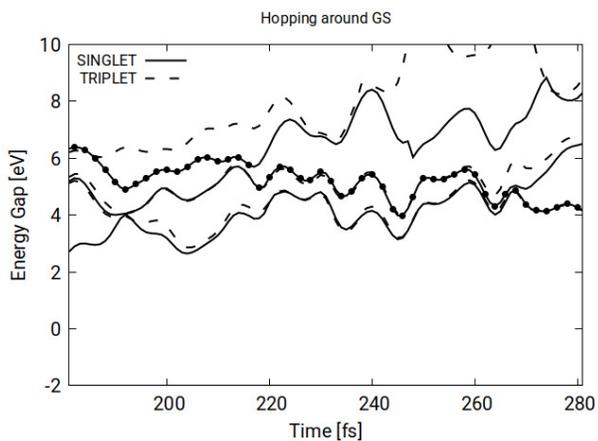
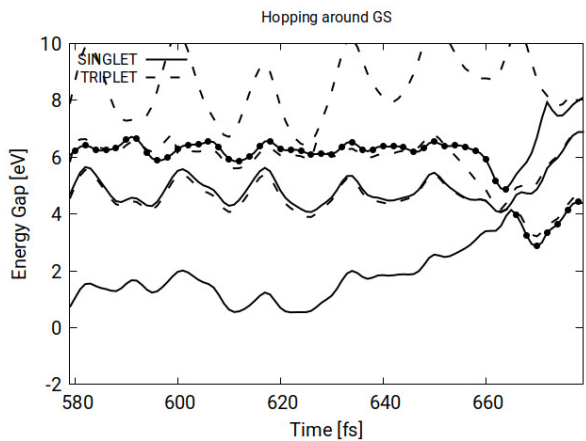
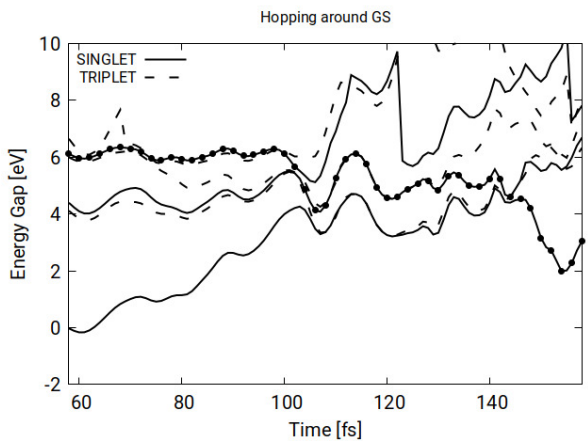

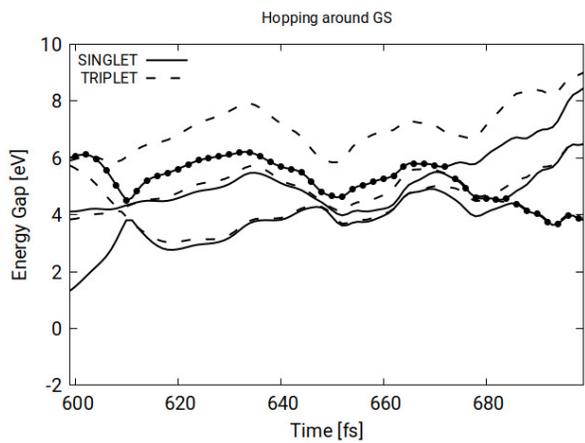
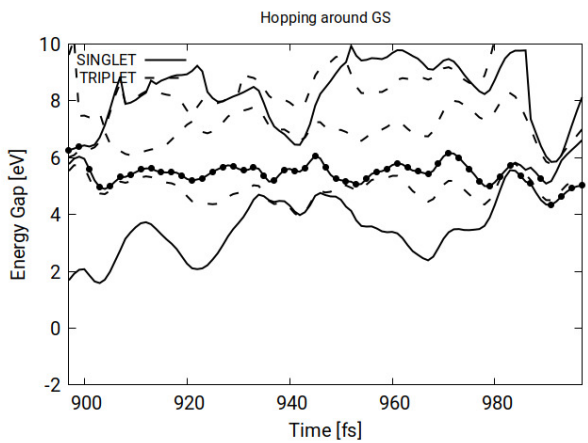
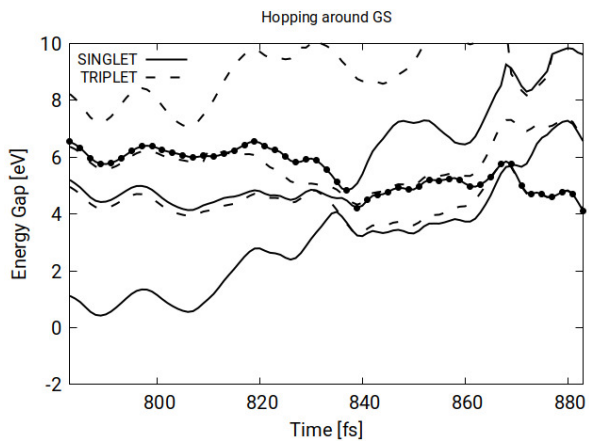
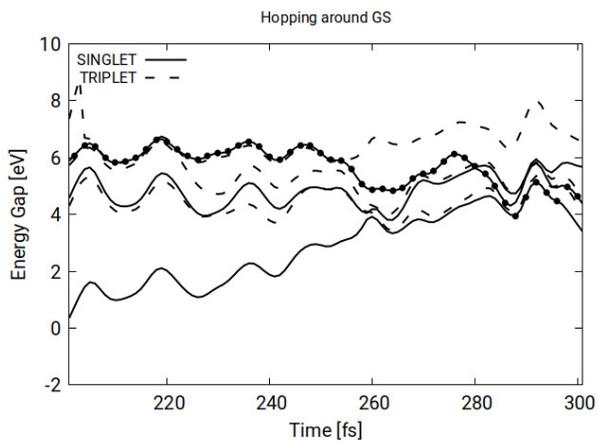
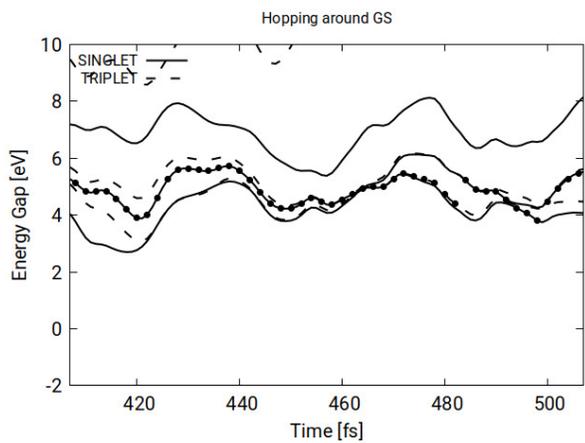
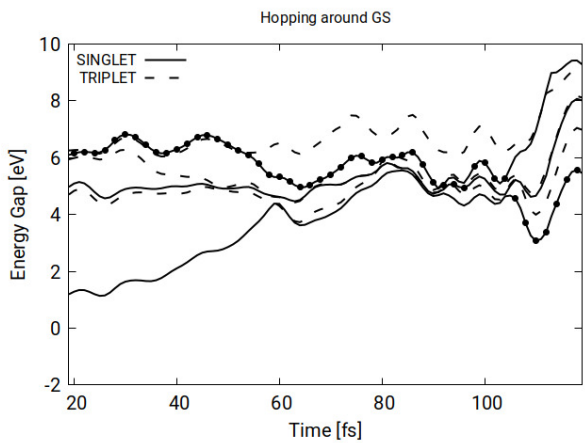
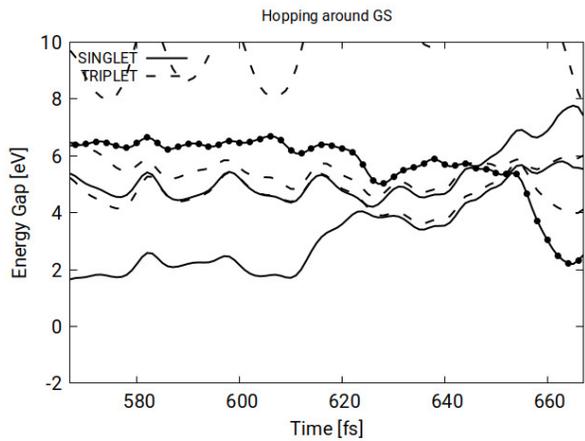
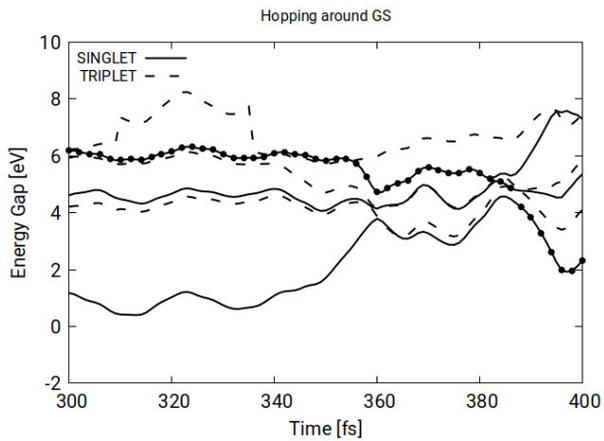

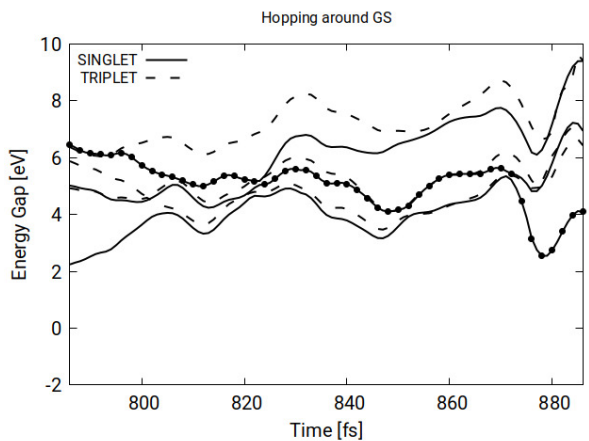
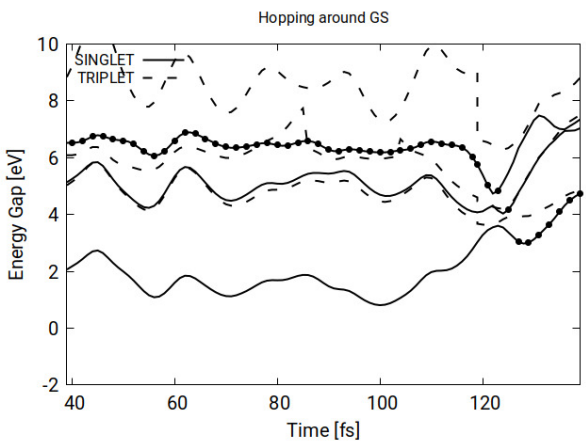
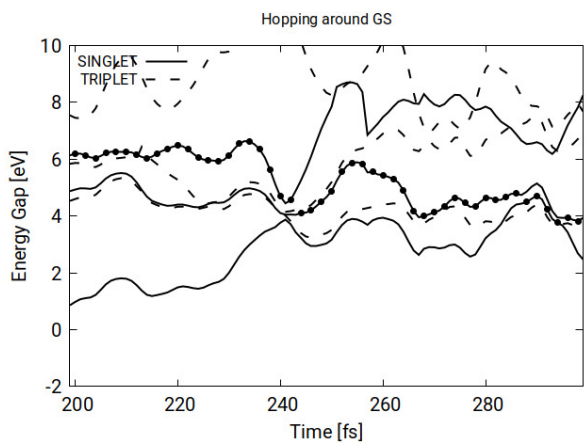
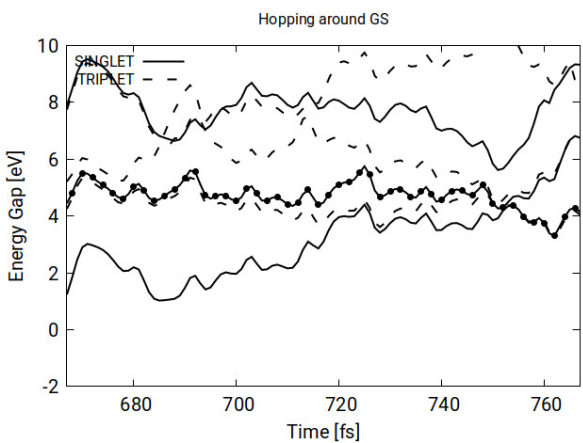
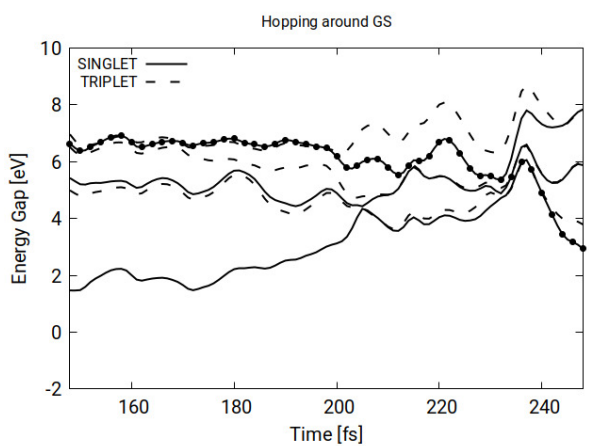
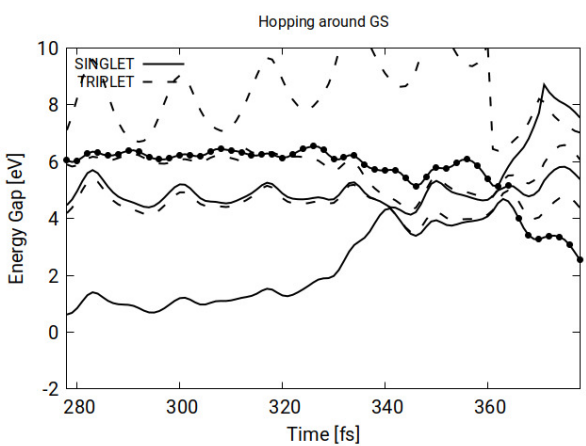
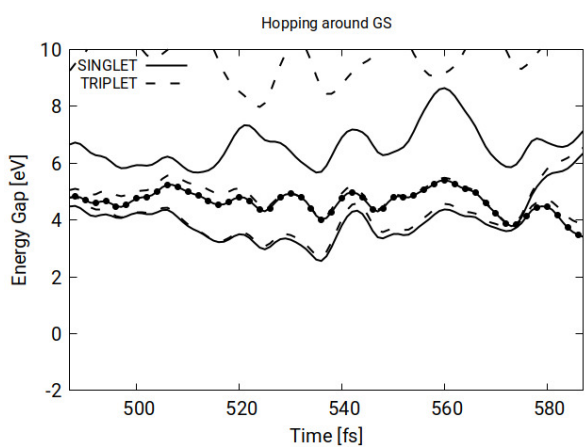